\documentclass[a4paper,twocolumn,11pt,accepted=2022-05-04]{quantumarticle}
\pdfoutput=1

\usepackage[utf8]{inputenc}
\usepackage[english]{babel}
\usepackage[T1]{fontenc}
\usepackage{amsmath}
\usepackage{hyperref}
\usepackage{braket}
\usepackage{multirow}
\usepackage{amsfonts}
\usepackage{amssymb}
\usepackage{leftidx}
\usepackage[shortlabels]{enumitem}
\usepackage{graphicx}
\usepackage[numbers,sort&compress]{natbib}

\begin{document}

\title{Time-Optimal Two- and Three-Qubit Gates for Rydberg Atoms}

\author{Sven Jandura}
\affiliation{University of Strasbourg and CNRS, CESQ and ISIS (UMR 7006), aQCess, 67000 Strasbourg, France}
\orcid{0000-0003-0282-7637}

\author{Guido Pupillo}
\affiliation{University of Strasbourg and CNRS, CESQ and ISIS (UMR 7006), aQCess, 67000 Strasbourg, France}

%Abbreviations
\newcommand{\omax}{\Omega_{\mathrm{max}}}
\newcommand{\beff}{B_{\mathrm{eff}}}
\renewcommand{\d}{\mathrm{d}}
\newcommand{\e}{\mathrm{e}}
\newcommand{\hc}{\mathrm{h.c.}}
\newcommand{\tr}{\mathrm{tr}}
\renewcommand{\phi}{\varphi}
\renewcommand{\Im}{\mathrm{Im}}
\renewcommand{\Re}{\mathrm{Re}}

\begin{abstract}
We identify time-optimal laser pulses to implement the controlled-Z gate and its three-qubit generalization, the C$_2$Z gate, for Rydberg atoms in the blockade regime. Pulses are optimized using a combination of numerical and semi-analytical quantum optimal control techniques that result in smooth Ans\"atze with just a  few variational parameters. For the CZ gate, the time-optimal implementation corresponds to a global laser pulse that does not require single-site addressability of the atoms, simplifying experimental implementation of the gate. We employ quantum optimal control techniques to mitigate errors arising due to the finite lifetime of Rydberg states and finite blockade strengths, while several other types of errors affecting the gates are directly mitigated by the short gate duration. For the considered error sources, we achieve theoretical gate fidelities compatible with error correction  using reasonable experimental parameters for CZ and C$_2$Z gates.
\end{abstract}

\maketitle

\section{Introduction}
The improvement of fidelities for two- and multi-qubit quantum gates is a main driver of research in the field of quantum computing: High-fidelity two-qubit gates are essential for the realization of deep quantum circuits in noisy near-term digital quantum devices~\cite{preskill_quantum_2018} as well as for the realization of fully-fledged fault tolerant quantum computers in the long term
~\cite{svore_noise_2007, fowler_surface_2012,spedalieri_latency_2009,lai_performance_2014}, while
high-fidelity $k$-qubit quantum gates ($k>2$) may drastically reduce the gate count for quantum algorithms and enable fault tolerant quantum computation schemes adapted to specific platforms~\cite{baker_exploiting_2021,cong_hardware-efficient_2021}.

\begin{figure*}[t]
    %\centering
    \includegraphics[width=\linewidth]{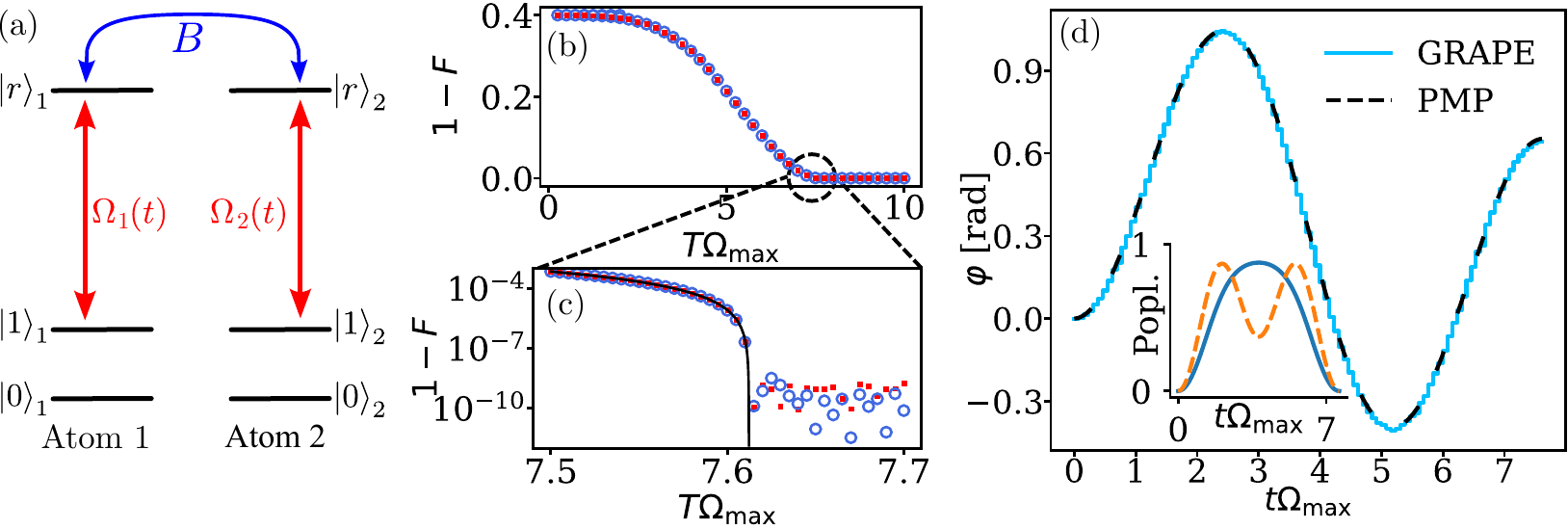}
    \caption{Time-optimal global CZ gate.  a) Level scheme used for $n=2$ atoms. Qubits are stored in electronic states of the atoms. The $\ket{1}_j$ state is coupled to the $\ket{r}_j$ state via a laser (red) with Rabi frequency $\Omega_j(t)$. Additionally there is a van der Waals interaction (blue) $B\ket{rr}_{12}\leftidx{_{12}}{\!\bra{rr}}{}$ if both atoms are in the Rydberg state. b) Smallest possible gate error $1-F$ found by GRAPE as a function of the dimensionless gate duration $T\omax$. Points shown by blue circles are obtained assuming a global laser, while points shown by red squares are obtained by assuming individual addressability of the atoms. At all $T$ the optimization in started with independent, random initial conditions. A piecewise constant Ansatz with 99 pieces was used for GRAPE, doubling the number of pieces decreases the gate error by at most $3\cdot 10^{-6}$, showing that 99 pieces are sufficient to capture the time-optimal pulse. c) Zoom-in at the gate errors, shown in log-scale, between $T\omax=7.5$ and $T\omax=7.7$. We start at $T\omax=7.7$ with a random initialization, the optimized pulse is then used as the initial guess at $T\omax = 7.695$ and so on.  The fit is $1-F = A(T_*-T)^2$ if $T<T_*$ and $1-F=0$ if $T \geq T_*$ with $T_* =7.612$ and $A=0.0544$.  Beyond the minimal pulse duration $T_*$ the gate error around $10^{-10}$ is given by the convergence condition of the optimization. d) Phase $\phi(t)$ of the time-optimal laser pulse at $T = T_*$ as found by GRAPE (solid line) and fitted with the PMP (dashed line). Inset: Population of $\ket{0r}$ (blue, solid line) and $\ket{W}$ (orange, dashed line) over time. }
    \label{fig:level_scheme_and_cz_panel}
\end{figure*}

Neutral atoms are among the leading technologies for advanced analog and digital quantum simulations and have recently emerged as a highly promising platform for quantum computing~\cite{bloch_many-body_2008,saffman_quantum_2010, browaeys_many-body_2020, henriet_quantum_2020, morgado_quantum_2021}.
Near defect-free arrays of alkali-metal and alkali-earth(-like) atoms can be routinely prepared at sub-millikelvin temperatures in optical tweezers and in optical lattices in arbitrary dimensions~\cite{endres_atom-by-atom_2016, barredo_atom-by-atom_2016, ohl_de_mello_defect-free_2019,barredo_synthetic_2018, schlosser_large-scale_2019}.
Fast two-qubit  gates can be achieved by encoding quantum information in the internal -- usually electronic ground -- states of individual atoms (or collective excitations of atomic ensembles~\cite{morgado_quantum_2021}) while strong interactions can be mediated by electronically highly excited Rydberg states, which can be controlled using laser fields [see Fig.~\ref{fig:level_scheme_and_cz_panel}(a)]. In the original scheme~\cite{jaksch_fast_2000}, the accumulation of relevant phase shifts in two-qubit quantum gates is facilitated by the so-called ``Rydberg blockade'' mechanism, whereby one laser-excited atom
shifts the Rydberg states of neighboring atoms out of resonance due to strong Rydberg-Rydberg interactions -- a process that can be readily generalized to $k$-qubit gates, in principle~\cite{muller_mesoscopic_2009, isenhower_multibit_2011}. Two-qubit entangling gates have been implemented in several experiments \cite{levine_parallel_2019, graham_rydberg-mediated_2019,picken_entanglement_2018,fu_high-fidelity_2022, martin_molmer-sorensen_2021,madjarov_high-fidelity_2020}, achieving fidelities up to 99.1\%. \cite{madjarov_high-fidelity_2020}.

A variety of error sources limit gate fidelities in experiments, including a finite Rydberg blockade strength, decay of the Rydberg state, scattering of an intermediate state in a two photon transition, laser phase noise, variations of the laser intensity with the position of the atom in the trap and Doppler shifts of the laser frequency due to thermal motion of the atoms \cite{graham_rydberg-mediated_2019, de_leseleuc_analysis_2018, zhang_fidelity_2012}. To mitigate the effects of these errors, many different improvements of the original protocol~\cite{jaksch_fast_2000} have been proposed based on adiabatic passage~\cite{rao_robust_2014,beterov_two-qubit_2016, mitra_robust_2020, saffman_symmetric_2020, beterov_application_2020, he_multiple-qubit_2021}, dark state mechanisms~\cite{petrosyan_high-fidelity_2017}, Rydberg Antiblockade~\cite{wu_unselective_2021, wu_resilient_2021, he_multiple-qubit_2021}, and many other approaches~\cite{theis_high-fidelity_2016, liu_optimized_2022,  shen_construction_2019, guo_optimized_2020, muller_mesoscopic_2009}. It is increasingly recognized that all these approaches can benefit from {\it
quantum optimal control methods} to improve both the speed and fidelities of the various quantum gates. Quantum optimal control methods can also provide additional solutions that differ qualitatively from any of the known basic gate schemes.

Quantum optimal control methods~\cite{glaser_training_2015} have been  successfully used on a variety of platforms, including superconducting qubits~\cite{egger_optimized_2014, kelly_optimal_2014, huang_optimal_2014, werninghaus_leakage_2021}, trapped ions~\cite{nebendahl_optimal_2009, choi_optimal_2014} and neutral atoms \cite{goerz_quantum_2011,muller_optimizing_2011, goerz_robustness_2014,omran_generation_2019, cui_optimal_2017, smith_quantum_2013, anderson_accurate_2015, lysne_small_2020}.
For neutral atoms, three properties of the laser pulses implementing a given gate are of particular interest: \textit{(i)} They should be time-optimal (i.e., as short as possible) because many errors mentioned above can be mitigated by short pulse durations; \textit{(ii)} They should ideally require only a global control laser addressing all atoms of the gate simultaneously -- so-called \emph{global} pulses --, instead of requiring single-site addressability, in order to simplify experiments;~\textit{(iii)} They should allow for executing three and more qubit gates {\it natively} (i.e., without decomoposing them into single- and two-qubit gates).
In line with these three requirements, a remarkable result in Ref.~\cite{levine_parallel_2019} shows that a global pulse for the Controlled-Z (CZ) gate exists, which is more then 30\% faster than the original, non-global proposal~\cite{jaksch_fast_2000}, and a global pulse exists for a Toffoli gate, albeit only with a moderate theoretically predicted fidelity of 97.6\%. Naturally the question arises if it is possible to find pulses that are even faster and achieve higher fidelities, in order to make both CZ and C$_2$Z gates fully compatible with error correction schemes.

In this work, we answer the fundamental question of identifying the fastest possible global pulse for a CZ gate and for its three-qubit generalization, the C$_2$Z gate, satisfying the requirements \textit{(i)-(iii)} above. While we assume the basic  level structure of Ref.~\cite{jaksch_fast_2000} (see Fig.~\ref{fig:level_scheme_and_cz_panel}(a)), our results provide original pulse schemes. Key results include: (i) For the CZ gate, we find that the  time-optimal global pulse corresponds to a pulse where the laser amplitude is kept constant, while the laser phase is varied continuously and smoothly in time. The resulting pulse is about 10\% faster than that of Ref.~\cite{levine_parallel_2019} -- which we thus find to be already an excellent pulse. Interestingly, the found global CZ gate remains time-optimal even when considering pulses making use of single-site addressability,  demonstrating that the latter is not necessary for two-qubit operations.
(ii) For the C$_2$Z gate, we find two qualitatively different time-optimal global pulses, which differ by less than 1\% in speed. Interestingly, both pulses are even faster than the pulse proposed in Ref.~\cite{isenhower_multibit_2011}, which requires single-site addressability. To our knowledge this is the first work to identify time-optimal pulses for C$_2$Z gates. (iii)
We demonstrate that the found time-optimal pulses can be adjusted to minimize errors arising from the decay of the Rydberg state or a finite blockade strength. (iv) All pulses for CZ and C$_2$Z gates correspond to smooth time evolutions of the laser phase that can be fully described by just a few variational parameters and (v) allow for reaching theoretical fidelities compatible with most error correction schemes.

The results in this work are obtained using two complementary quantum optimal control techniques, namely Gradient Ascent Pulse Engineering (GRAPE) and Pontryagins Maximum Principle (PMP), which we combine in a novel way: The time-optimal pulses are first found using GRAPE, which requires optimization over several hundreds of parameters to describe the pulses. The pulses can then be fitted by the solution of a simple differential equation, derived using the PMP from the condition of time-optimality. In this way, the number for parameters needed to describe the pulses is reduced to only 4 and 6 for the CZ and the C$_2$Z gate, respectively.

The paper is structured as follows: In Sec.~\ref{sec:theorectical_tools} we introduce the theoretical tools used in this work. In particular, in Sec.~\ref{subsec:hamiltonians_and_pulses} we introduce the level scheme and the Hamiltonian, and define the averaged fidelity as a quality measure for the implementation of a gate. We also discuss how complex conjugation or time-reversal of a laser pulse affect the implemented gate and how calculations can be simplified for global pulses. In Secs.~\ref{subsec:grape} and \ref{subsec:pmp} we give a brief introduction to GRAPE and the PMP, respectively. In Sec.~\ref{sec:time_optimal_gates_at_infinite_blockade_strengths} we  use GRAPE to find the time-optimal global pulses that implement a CZ and a C$_2$Z gate in the limit of an infinitely large blockade strength. In Sec.~\ref{sec:semi-analytical_description_using_the_pmp} we then use the PMP to give a semi-analytical description of the time-optimal pulses found in Sec.~\ref{sec:time_optimal_gates_at_infinite_blockade_strengths}. In Sec.~\ref{sec:minimizing_the_decay_of_the_rydberg_state} we change the optimization objective and use GRAPE to identify pulses that are the most robust against decay of the Rydberg state. In Sec.~\ref{sec:gates_at_finite_blockade_strength} we discuss how the time-optimal gates from Sec.~\ref{sec:time_optimal_gates_at_infinite_blockade_strengths} can be modified to compensate for a finite blockade strength. For this, we consider two approaches: In Sec.~\ref{subsec:fixed_blockade_strength} we assume a fixed and known blockade strength and find pulses that implement a CZ or C$_2$Z gate exactly only at this specific blockade strength, while in Sec.~\ref{subsec:variable_blockade_strength} we find pulses that implement a CZ or C$_2$Z gate exactly only at infinite blockade strength, but whose error increases as slowly as possible if the blockade strength is decreased. In Sec.~\ref{sec:infidelity_for_a_specific_setup} we calculate the fidelity for the time-optimal pulses from Sec.~\ref{sec:time_optimal_gates_at_infinite_blockade_strengths} for specific experimental parameters. Section~\ref{sec:conclusion} presents the conclusions.

All pulses found in this work are available at Ref.~\cite{pulses_figshare}.

\section{Theoretical Tools}
\label{sec:theorectical_tools}

In this section we introduce the theoretical tools that are used to derive optimal pulses in the subsequent sections.  Section~\ref{subsec:hamiltonians_and_pulses} introduces the Hamiltonian and presents the basic properties of pulses implementing a quantum gate.  Section~\ref{subsec:quantum_optimal_control_methods} gives a brief introduction to GRAPE, followed by an introduction to the application of the PMP method to time-optimal control.

\subsection{Hamiltonians and Pulses}
\label{subsec:hamiltonians_and_pulses}
In Sec.~\ref{subsubsec:general_hamiltonian} we first introduce the level scheme and the corresponding Hamiltonian that are used in remainder of the work. In Sec.~\ref{subsubsec:phase_gates_and_infidelity} we introduce phase gates, of which the CZ gate and the C$_2$Z gate are two examples, and show how the gate error of a pulse aiming to implement a phase gate is calculated. In Sec~\ref{subsubsec:symmetry_operations} we discuss how complex conjugation and/or time reversal of a pulse affect the implemented gate and show that in general there is not only one time-optimal pulse for a given gate, but several pulses which are related by symmetry operations. Finally, in Sec.~\ref{subsubsec:global_pulses} we focus on global pulses and show how some calculations simplify for them.

\subsubsection{General Hamiltonian}
\label{subsubsec:general_hamiltonian}
Consider $n$ atoms treated as three-level systems: A qubit is stored in states $\ket{0}$ and $\ket{1}$, which can be taken to be hyperfine states of the ground state manifold. Additionally, a Rydberg state $\ket{r}$ is used to mediate the interactions between the qubits. This work focuses on the cases $n=2$ and $n=3$ to describe two- and three-qubit gates.

The level scheme for $n=2$ is shown in Fig.~\ref{fig:level_scheme_and_cz_panel}(a).  The states $\ket{1}_j$ and $\ket{r}_j$ of the $j$-th atom are coupled by a laser with time-dependent Rabi frequency $\Omega_j(t) = |\Omega_j(t)|e^{i\phi_j(t)}$. The Rabi frequency $\Omega_j(t)$ is taken to be complex, encoding both the amplitude $|\Omega_j(t)|$ and the phase $\phi_j(t)$ of the laser. For this reason, no additional detuning of the laser is included, since the laser detuning $\Delta_j(t)$ and the laser phase are related through $\Delta_j = \d \phi_j/\d t$. In any experiment, the maximal achievable Rabi frequency $|\Omega_j|$ is limited by the laser power and waist diameter. To include this into the model, a maximum Rabi frequency $\omax$ is introduced and only pulses with $|\Omega_j(t)| \leq \omax$ are considered for the rest of the paper.

The inter-atomic interaction term $B_{jk}\ket{rr}_{jk}\leftidx{_{jk}}{\!\bra{rr}}{}$ shifts the energy of the atoms $j$ and $k$ that are prepared in the Rydberg state, with $B_{jk}$ called the \emph{blockade strength}. This results in the Hamiltonian (with $\hbar = 1$)
\begin{align}
    \label{eq:general_hamiltonian}
     H(t) =& \sum_{i=j}^n \frac{\Omega_j(t)}{2}\ket{1}_j\leftidx{_j}{\!\bra{r}}{} + \hc \\ \nonumber
          +& \sum_{j<k} B_{jk} \ket{rr}_{jk}\leftidx{_{jk}}{\!\bra{rr}}{}.
\end{align}
The Hamiltonian~\eqref{eq:general_hamiltonian} ignores any Rydberg-laser induced light shifts on $\ket{0}$, $\ket{1}$ and $\ket{r}$.  These light shifts can be cancelled by adjusting the laser frequency and by applying additional single qubit rotations around the $z$ axis after the pulse.

In previous studies, blockade strengths up to $B_{jk}/{2\pi}=3$GHz have been considered \cite{saffman_symmetric_2020, theis_high-fidelity_2016}. In Appendix~\ref{appendix:estimation_of_effective_blockade_strength} we provide a concrete example of how a blockade strength of $B_{jk}/2\pi = 180$MHz can be achieved in experiments, taking fully into account the dipole-dipole interaction between all relevant Rydberg states. The  corresponding gate pulse errors for blockade strengths between $100\mathrm{MHz} < B_{jk}/2\pi < 3\mathrm{GHz}$  are calculated in Sec.~\ref{sec:infidelity_for_a_specific_setup}. We note that the coupling between the electronic and the motional degree of freedom \cite{robicheaux_photon-recoil_2021} is not treated explicitly in Eq.~\eqref{eq:general_hamiltonian}. %this work.
However, we expect that the gate errors arising due to this coupling are mitigated by the short gate durations of the time-optimal pulses.

For $q \in \{0,1\}^n$ we denote by $P_q$ the projector onto the space spanned by all states such that atom $j$ is in state $\ket{0}$ iff $q_j=0$. For example, at $n=2$ we have  $P_{00} = \ket{00}\bra{00}$, $P_{01}=\ket{01}\bra{01} + \ket{0r}\bra{0r}$ and $P_{11} = \ket{11}\bra{11} +\ket{1r}\bra{1r} + \ket{r1}\bra{r1} + \ket{rr}\bra{rr}$ (Here and in the following, a bra or a ket vector without a subscript means that this vector describes the state of \emph{all}  atoms, e.g. $\ket{00}=\ket{00}_{12}$). Since the lasers only couple $\ket{1}_j$ to $\ket{r}_j$, $H$ is block-diagonal with respect to the $P_q$, i.e.
\begin{equation}
    H = \sum_q H_q \qquad H_q = P_q H P_q.
    \label{eq:H_q_definition}
\end{equation}
This block-diagonality allows for solving the time evolution of a computational basis state $\ket{q}$ by solving the Schrödinger equation for $H_q$ instead of $H$, which reduces the dimension of the relevant Hilbert space. This will simplify the use of GRAPE and PMP methods in Secs.~\ref{sec:time_optimal_gates_at_infinite_blockade_strengths} and \ref{sec:semi-analytical_description_using_the_pmp}.

\subsubsection{Phase Gates and Gate Error}
\label{subsubsec:phase_gates_and_infidelity}
Because any quantum gate has to map a state in the computational subspace $\mathcal{C} = \mathrm{span}\{\ket{q}| q \in \{0,1\}^n\}$ back to $\mathcal{C}$ and $H$ is block diagonal with respect to the $P_q$, the only quantum gates that can be implemented with Eq.~\eqref{eq:general_hamiltonian} are phase gates.

Given laser pulses $\Omega_1,...,\Omega_n$ of duration $T$, we denote by $U(t) = \tau\exp\left(-i \int_0^t H(t')\d t'\right)$ the time evolution operator. The pulse is said to implement a $(\xi_q)_{q \in \{0,1\}^n}$ phase gate if
\begin{equation}
    U(T)\ket{q} = e^{i\xi_q}\ket{q}.
\end{equation}
Note that since $H_{0...0}=0$ only phase gates with $\xi_{0...0}=0$ can be implemented. A phase gate on $n=2$ qubits is, up to single qubit gates, a CZ gate if $\xi_{11}-\xi_{01}-\xi_{10} = \pi$. A phase gate on $n=3$ qubits is, up to single qubit gates, a C$_2$Z gate if $\xi_{111} - \xi_{001}-\xi_{010}-\xi_{100} = \pi$ and $\xi_{011} = \xi_{001}+\xi_{010}$ and all permutations of the second equation hold.

The averaged fidelity $F$ of a pulse aiming to implement a $(\xi_q)_q$ phase gate is defined by
\begin{equation}
    F = \int_\mathcal{C} \d \psi \left| \left\langle \psi | U_0^\dag U(T)|\psi\right\rangle \right|^2
\end{equation}
where $U_0 = \sum_q e^{i\xi_q}\ket{q}\bra{q}$ is the desired phase gate and the integral is taken over all normalized states in $\mathcal{C}$ and with respect to the Haar measure on the unit sphere $\mathbb{S}^{2^n-1}$. The integral can be evaluated to \cite{pedersen_fidelity_2007}
\begin{align}
 \label{eq:fidelity_time_evolution_operator}
    F = \frac{1}{2^n(2^n+1)}\Big( & \left|\sum_q e^{-i\xi_q}\braket{q|U(T)|q}\right|^2 \\ \nonumber
    &+ \sum_q \left|\braket{q|U(T)|q}\right|^2\Big).
\end{align}
Interestingly, the fidelity can be found by propagating a single (unnormalized) state -- namely $\ket{\psi(0)} = \sum_q \ket{q}$ -- under $H(t)$, since by the bock-diagonality of $H$ it holds that $\braket{q|U(T)|q} =  \braket{q|\psi(T)}$. The so called \emph{gate error} $1-F$ is used in the following as a quality measure for a pulse aiming to implement a $(\xi_q)_q$ phase gate. In Sec.~\ref{sec:time_optimal_gates_at_infinite_blockade_strengths}, GRAPE will be used to time-optimally steer $\ket{\psi(0)}$ to $U_0\ket{\psi(0)}$ by minimizing the gate error, which then provides a time-optimal pulse to implement the phase gate by considering the evolution of a single state only.

For the specific case of a CZ gate as a phase gate the Bell state fidelity is commonly used instead of the averaged fidelity \cite{robicheaux_photon-recoil_2021, graham_rydberg-mediated_2019, levine_parallel_2019, theis_high-fidelity_2016} . The Bell state fidelity measures the quality of a pulse by the fidelity between the state obtained by applying the pulse to $\ket{++}=(\ket{00}+\ket{01}+\ket{10}+\ket{11})/2$ and the desired output state $CZ\ket{++}$. It is thus given by $F_{\mathrm{bell}} = |\braket{00|\psi_{00}} + \braket{01|\psi_{01}} + \braket{10|\psi_{10}}-\braket{11|\psi_{11}}|^2/16$, which can be generalized to arbitrary phase gates as
\begin{equation}
\label{eq:bell_state_fidelity}
F_{\mathrm{bell}} = 4^{-n} \left|\sum_q e^{-i\xi_q}\braket{q|U(T)|q}\right|^2
\end{equation}
Comparing Eq.~\eqref{eq:fidelity_time_evolution_operator} and Eq.~\eqref{eq:bell_state_fidelity} shows that the averaged fidelity puts more emphasis on errors leading outside of the computational subspace than the Bell state fidelity. In this work we use the averaged fidelity. We expect that the same pulses can be obtained by optimizing the Bell state fidelity instead.

\subsubsection{Symmetry Operations}
\label{subsubsec:symmetry_operations}
The time-optimal pulse implementing a phase gate is not necessarily unique, as other pulses implementing the same or a related gate can be obtained by symmetry operations. Here, three of these symmetry operations are discussed: Phase shifts, time-reversal,  complex conjugation.

Given a pulse $\Omega_1, ..., \Omega_n$ of duration $T$ implementing a $(\xi_q)_q$ phase gate for a blockade strength $B_{jk}$, several other pulses $\tilde{\Omega}_1, ..., \tilde{\Omega}_n$ with the same duration $T$ can be constructed that implement a $(\tilde{\xi}_q)_q$ phase gate using the following symmetry operations
\begin{enumerate}[a)]
    \item \emph{Phase shifts}:  $\tilde{\Omega}_j(t) = e^{i\alpha_j}\Omega_j(t)$, $\tilde{B}_{jk} = B_{jk}$ and $\tilde{\xi}_q = \xi_q$, for constants $\alpha_1, ..., \alpha_n$. To prove this statement, we denote here and in the following by $\tilde{H}$ the Hamiltonian given by the pulses $\tilde{\Omega}_1,..., \tilde{\Omega}_n$ and blockade strengths $\tilde{B}_{ij}$ according to Eq. \eqref{eq:general_hamiltonian}. Further, let $\tilde{U}(t)$ be the time evolution operator under $\tilde{H}$. Then with the basis change $V = \bigotimes_j (\ket{0}\bra{0}+\ket{1}\bra{1}+e^{-i\alpha_j}\ket{r}\bra{r})$ we have $\tilde{H} = V HV^\dag$.  Hence $\tilde{U}(T) = VU(T)V^\dag$, thus also $\braket{q|\tilde{U}(T)|q} = \braket{q|U(T)|q} = e^{i\xi_q}$.
     \label{item:phaseshift}
    \item \emph{Time reversal}:  $\tilde{\Omega}_i(t) = \Omega_i(T-t)$, $\tilde{B}_{ij}=-B_{ij}$ and $\tilde{\xi}_q = - \xi_q$. To prove that, we first show that another symmetry is given by $\tilde{\Omega}_i(t) = -\Omega_i(T-t)$, $\tilde{B}_{ij} = -B_{ij}$ and $\tilde{\xi}_q = - \xi_q$. To see this, note that $\tilde{H}(t) = -H(T-t)$, so the time evolution operator under $\tilde{H}$ is $\tilde{U}(t) = U(T-t)U(T)^\dag$. Hence $\braket{q|\tilde{U}(T)|q} = \braket{q|U(T)^\dag|q} = e^{-i\xi_q}$. Together with the symmetry under phase shifts, symmetry under time reversal follows.
    \label{item:time_reversal}
    \item \emph{Complex conjugation}: $\tilde{\Omega}_i(t) = \Omega_i(t)^*$, $\tilde{B}_{ij}=-B_{ij}$ and $\tilde{\xi}_q = - \xi_q$. To prove this, we note that joint time reversal and complex conjugation is a symmetry, given by $\tilde{\Omega}_i(t) = \Omega_i(T-t)^*$, $\tilde{B}_{ij} = B_{ij}$ and $\tilde{\xi}_q = \xi_q$. To see this,  note that $\tilde{H}(t) = H(T-t)^*$, so the time-evolution operator under $\tilde{H}$ is $\tilde{U}(t) = U(T-t)^*(U(T)^*)^\dag$. Hence $\braket{q|\tilde{U}(T)|q} = \braket{q|(U(T)^*)^\dag|q} = e^{i\xi_q}$. Together with symmetry under time reversal, symmetry under complex conjugation follows.
    \label{item:complex_conjugation}
\end{enumerate}

These three symmetry operations show that when using numerical methods to find the time-optimal pulse for a gate, several solutions are to be expected. Firstly, two time-optimal pulses can differ by an arbitrary constant phase. This degree of freedom can be eliminated by restricting the discussion to pulses where $\Omega_j(0)$ is real and positive, as we will do from now on. Even with this restriction there are in general two different time-optimal pulses for a given phase gate, related by joint complex conjugation and time reversal. In the special case of $B=\infty$ and for phase gates where all $\xi_q$ are real, which is studied in Sec.~\ref{sec:time_optimal_gates_at_infinite_blockade_strengths}, there are in general even four different time-optimal pulses, because also individual time reversal or complex conjugation give pulses implementing the same gate. As will be apparent in Sec.~\ref{sec:time_optimal_gates_at_infinite_blockade_strengths}, sometimes, but not always, time-optimal pulses are invariant under joint time-reversal and complex conjugation, reducing the number of distinct time-optimal pulses to two.

\subsubsection{Global Pulses}
\label{subsubsec:global_pulses}
This work focuses primarily on so called \emph{global} pulses, where $\Omega_1 = \Omega_2 = ... = \Omega_n =: \Omega$. In fact, except in Sec.~\ref{subsec:non_parallel_cz_gate}, all considered pulses will be global. Global pulses can be implemented with just a single global laser addressing all atoms at the same time, no single-site addressability being needed. They are thus expected to be easier to realize experimentally. When working with global pulses it will be  further assumed (except briefly in Sec.~\ref{sec:gates_at_finite_blockade_strength}) that all $B_{jk}$ are equal to the same blockade strength $B$.

The theoretical treatment using global pulses is simplified compared to the general case in two different ways: Firstly, now it holds that $\braket{q|\psi(T)} = \braket{q'|\psi(T)}$ if $q$ and $q'$ have the same number of 1s. As a consequence, only phase gates with $\xi_q = \xi_{q'}$ can be implemented. For $n=2$ the whole gate is then determined by the evolution of $\ket{\psi(0)} = \ket{01} + \ket{11}$ under $H_{01}+H_{11}$, and the fidelity is given by

\begin{align}
    \label{eq:fidelity_cz}
    F = \frac{1}{20} & \Big( \left|1+2a_{01}+a_{11}\right|^2 \\ \nonumber
        & +1+ 2|a_{01}|^2+|a_{11}|^2\Big)
\end{align}
with $a_q = e^{-i\xi_q} \braket{q|\psi(T)}$.
Similarly, for $n=3$ the whole gate is determined by the evolution of $\ket{\psi(0)} = \ket{001}+\ket{011}+\ket{111}$ under $H_{001}+H_{011}+H_{111}$, with
\begin{align}
    \label{eq:fidelity_ccz}
    F = \frac{1}{72} & \Big( \left| 1 + 3a_{001} + 3a_{011} + a_{111}\right|^2 \\ \nonumber
    &+ 1 + 3|a_{001}|^2 + 3|a_{011}|^2 + |a_{111}|^2\Big)
\end{align}

The second simplification due to global pulses  is a simplification of Hamiltonians $H_q$ if $q$ contains two or more 1s. For $n=2$ we have
\begin{align}
    H_{11}=& \frac{\Omega}{2}\Big( \ket{11}\bra{1r} + \ket{11}\bra{r1} + \ket{1r}\bra{rr} \\ \nonumber
    &+ \ket{r1}\bra{rr}\Big) + \hc + B\ket{rr}\bra{rr}  \\ \nonumber
    =& \frac{\sqrt{2}\Omega}{2}\Big( \ket{11}\bra{W} + \ket{W}\bra{11} \Big)+\hc \\ \nonumber
    &+ B\ket{rr}\bra{rr}
\end{align}
with $\ket{W} = (\ket{1r}+\ket{r1})/\sqrt{2}$. The rank of $H_{11}$ thus reduces from generally 4 to 3 in the global case. For $n=3$ we make the analogous simplification
\begin{equation}
    H_{011} = \ket{0}\bra{0} \otimes H_{11}.
\end{equation}
To simplify $H_{111}$, we note that the only states that can be accessed from $\ket{111}$ with a global pulse are $\ket{W_1} = (\ket{11r} + \ket{1r1} + \ket{r11})/\sqrt{3}$, $\ket{W_2} = (\ket{1rr}+\ket{r1r} + \ket{rr1})/\sqrt{3}$ and $\ket{rrr}$. It is thus sufficient to consider just the projection of $H_{111}$ onto the symmetric subspace $\mathrm{span}\{\ket{111}, \ket{W_1}, \ket{W_2}, \ket{rrr}\}$, given by
\begin{align}
    H^{\mathrm{sym}}_{111} &= \frac{\sqrt{3}\Omega}{2}\ket{111}\bra{W_1} + \Omega \ket{W_1}\bra{W_2} \nonumber \\
    &+ \frac{\sqrt{3}\Omega}{2} \ket{W_2}\bra{111} + \hc \nonumber \\
    &+ B \ket{W_2}\bra{W_2} + 3B\ket{rrr}\bra{rrr}.
\end{align}
This reduces the rank of $H_{111}$ from generally 8 to 4 in the global case.

This discussion concludes the presentation of the basic properties of pulses implementing phase gates, and of global pulses in particular. After a brief introduction to GRAPE and the PMP in the following subsection, the theoretical discussion from this section will be used in Secs.~\ref{sec:time_optimal_gates_at_infinite_blockade_strengths} and \ref{sec:semi-analytical_description_using_the_pmp} to identify the time-optimal pulses for the CZ gate and the C$_2$Z gate.

\subsection{Quantum Optimal Control Methods}
In this section we give a brief introduction to GRAPE and the PMP, the two quantum optimal control methods used in this work.
\label{subsec:quantum_optimal_control_methods}
\subsubsection{GRAPE Algorithm}
\label{subsec:grape}
Gradient Ascent Pulse Engineering (GRAPE) is a quantum optimal control technique based on gradient descent. It was originally developed to design pulse sequences for NMR spectroscopy, and has been used successfully to speed up quantum gates and to increase the fidelity of pulses in the presence of imperfections \cite{khaneja_optimal_2005, garon_time-optimal_2013,riaz_optimal_2019, wilhelm_introduction_2020}. In the context of neutral atoms GRAPE has been used to design arbitrary $SU(16)$ gates in the hyperfine levels of $~^{133}$CS \cite{smith_quantum_2013, anderson_accurate_2015, lysne_small_2020}.

Given an initial state $\ket{\psi(0)}$, a pulse duration $T$ and a Hamiltonian $H$ depending on controls $u(t)$, GRAPE can be used to find the optimal controls $u(t)$, minimizing $J(\ket{\psi(T)})$, for an arbitrary objective function $J$. For this, a piecewise constant Ansatz for $u(t)$ is made, i.e. $u(t) = u_j$ if $t \in [j\Delta t, (j+1)\Delta t]$ and $\Delta t = T/N$, with $N$ the number of pieces. Typically, several hundreds of pieces are used. Minimizing $J(\ket{\psi(T)})$ becomes an optimization problem over $u_0,...,u_{N-1}$. GRAPE provides a fast way to calculate all derivatives $\partial J /\partial u_j$, which then allows for the efficient use of a gradient descent algorithm to find a local minimum of the cost function. In our implementation, we use the Broyden–Fletcher–Goldfarb–Shanno algorithm \cite{nocedal_numerical_2006, jones_scipy_2001} as the gradient descent method.

\subsubsection{Pontryagin's Maximum Principle for Time-Optimal Control}
\label{subsec:pmp}
Pontryagin's Maximum principle (PMP) \cite{pontryagin_mathematical_1986,lee_foundations_1986, boscain_introduction_2021} is an optimal control technique that has been used successfully to find time-optimal~\cite{bao_optimal_2018, lin_application_2019, lin_time-optimal_2020, garon_time-optimal_2013} or robust~\cite{van_damme_robust_2017} quantum gates, and to optimize variational quantum algorithms~\cite{yang_optimizing_2017}. The PMP is an analytic principle that gives a set of necessary conditions that optimal pulses have to satisfy, thereby allowing to reduce the infinite dimensional control landscape to a low-dimensional space \cite{van_damme_robust_2017}. In contrast to GRAPE, which usually uses several hundreds of variational parameters, the PMP can therefore describe optimal pulses with just a handful of parameters. On the downside, the PMP does not provide a universal algorithm to find the parameters leading to the optimal pulse. In Sec.~\ref{sec:semi-analytical_description_using_the_pmp} we will combine GRAPE and the PMP to describe the pulses found by GRAPE using the PMP with just 4 parameters for the CZ and 6 parameters for the C$_2$Z gate.

In this section we give a formulation of the PMP and apply it to the problem of time-optimal pulses on Rydberg atoms. We start by formulating the PMP for the specific case of time-optimal control problems~\cite{lee_foundations_1986}: Suppose one wants to steer a real vector $x \in \mathbb{R}^n$ from an initial point $x_i$ to a final point $x_f$ using the differential equation $\dot{x} = f(x,u)$, where $u(t)$ describe the available controls (i.e., the Rabi frequencies $\Omega_1,...,\Omega_n$ of the lasers in our case). Later, $x$ will correspond to the quantum state and $f$ will be given by Schrödinger's equation. We call a triplet $(T, x(\cdot), u(\cdot))$ a \emph{controlled trajectory} if $x(0)=x_i$, $x(T)=x_f$ and $\dot{x}(t) = f(x(t), u(t))$. The PMP gives necessary conditions on the controlled trajectory $(T, x, u)$ that minimizes $\int_0^T g(x(t), u(t)) \d t$, where $g$ is an arbitrary cost function.

By focusing on time-optimal control problems, we can fix $g(x,u)=1$. In this case, the PMP then states that for a given time-optimal controlled trajectory $(T, x, u)$ there exist so-called \emph{costates} $\lambda(\cdot)$ with values in $\mathbb{R}^n\setminus \{0\}$ such that at each time the following equations are satisfied
\begin{equation}
    \dot{\lambda} = - \left\langle \lambda, \nabla_x f(x,u)\right\rangle
    \label{eq:pmp1}
\end{equation}
and
\begin{equation}
    \left\langle \lambda(t), f(x(t),u(t))\right\rangle = \sup_{u'} \left\langle \lambda(t), f(x(t),u')\right\rangle,
    \label{eq:pmp2}
\end{equation}
where $\left\langle \cdot, \cdot \right\rangle$ denotes the scalar product. Further, if the $x$ are restricted to a sub\-mani\-fold $M$ of $\mathbb{R}^n$, the $\lambda(t)$ can be restricted to the tangent space of $M$ at $x(t)$ \cite{chang_simple_2011}. If the supremum in Eq.~\eqref{eq:pmp2} is only achieved by a single value $u'$ of the controls, the PMP allows to reduce the search for the time-optimal trajectory to the search over the initial costates $\lambda(0)$, which determine the full trajectory via Eqs.~\eqref{eq:pmp1} and  \eqref{eq:pmp2}.

In this work, we directly apply the PMP to the Schrödinger equation $\ket{\dot{\psi}} = -iH(u)\ket{\psi}$. This is done by decomposing the state $\ket{\psi} = \ket{\psi_R}+i\ket{\psi_I}$ and the Hamiltonian $H=H_R+iH_I$ into  real and imaginary parts~\cite{lin_application_2019, lin_time-optimal_2020}, which results in two real costates $\ket{\chi_R}$ and $\ket{\chi_I}$. By combining the two real costates with a complex one as $\ket{\chi} = \ket{\chi_R}+i\ket{\chi_I}$, Eq.~\eqref{eq:pmp1} becomes the Schrödinger equation $\ket{\dot{\chi}} =-iH\ket{\chi}$ for the costates, while Eq.~\eqref{eq:pmp2} reads \cite{lin_time-optimal_2020}
%\begin{equation}
%    \Im(\braket{\chi(t)|H(u(t))|\psi(t)}) = \sup_{u'} \Im(\braket{\chi(t)|H(u')|\psi(t)}).
%    \label{eq:pmp_maximization_seq}
%\end{equation}
\begin{align}
    \label{eq:pmp_maximization_seq}
     &\Im(\braket{\chi(t)|H(u(t))|\psi(t)}) \\ \nonumber
     &= \sup_{u'} \Im(\braket{\chi(t)|H(u')|\psi(t)}).
\end{align}

The key point is that, as long as the supremum is achieved at a unique value of the controls, the initial states and costates completely determine the optimal trajectory.

Applied to time-optimal phase gates the PMP states that there are costates $\ket{\chi_q(t)}$ evolving under the Schrödinger equation $\ket{\dot{\chi}_q} = -iH_q\ket{\chi_q}$ such that
\begin{align}
    \label{eq:pmp_phase_gates}
    &\Im\left(\sum_q \braket{\chi_q(t)| H_q(u(t)) | \psi_q(t)}\right) \\ \nonumber
    &= \sup_{u'}\Im\left(\sum_q \braket{\chi_q(t)| H_q(u') | \psi_q(t)}\right)
\end{align}
where the sum can be either over all $q \in \{0,1\}^n$, or for global pulses with $n=2$ or $n=3$ only over $q \in \{01, 11\}$ or $q \in \{001, 011, 111\}$, respectively.

In Sec.~\ref{sec:semi-analytical_description_using_the_pmp} we will use the PMP to reproduce the results found by GRAPE by extracting the initial costates from the pulses found by GRAPE. This allows us to describe the time-optimal pulses just through the initial costates, which correspond to 4 parameters for the CZ gate and 6 parameters for the C$_2$Z gate.

\section{Time-Optimal Gates at Infinite Blockade Strength ($B=\infty$)}
\label{sec:time_optimal_gates_at_infinite_blockade_strengths}
In this section we use GRAPE to find the time-optimal pulses in the limit $B=\infty$. In this limit states with more than one atom in the Rydberg state $\ket{r}$ cannot be populated, because the lasers coupling them to states with one atom in the Rydberg state are infinitely far detuned.  We start by considering the CZ gate in Sec.~\ref{subsec:parallel_cz_gate} and find that the time-optimal global pulse is approximately 10\% faster than the pulse of Ref.~\cite{levine_parallel_2019}. In Sec.~\ref{subsec:non_parallel_cz_gate} we then show that allowing for individual adressability for the atoms brings no speedup for the CZ gate, which is one of the main results of the work. Finally, in Sec.~\ref{subsec:parallel_ccz_gate} we find the time-optimal global pulse for the three-qubit C$_2$Z gate, and a second, slightly slower global pulse that also implements a C$_2$Z gate. All found pulses are smooth functions of the parameters.

\subsection{Global CZ Gate}
\label{subsec:parallel_cz_gate}
In this section we apply GRAPE to find the time-optimal global pulse that implements a CZ gate in the setup of Fig.~\ref{fig:level_scheme_and_cz_panel}(a) with $B=\infty$. We say that a pulse implements a CZ gate if it implements a $(\xi_q)_q$ phase gate with $\xi_{00} = 0$, $\xi_{01} = \xi_{10} = \theta$ and $\xi_{11} = 2\theta +\pi$ for some single qubit phase $\theta$. By additional single qubit gates compensating the phase of $\theta$ gained by $\ket{01}$ and $\ket{10}$, a CZ gate with $\ket{xy} \mapsto (-1)^{xy}\ket{xy}$ can be achieved.

As discussed in Sec.~\ref{subsec:hamiltonians_and_pulses}, the action of a global pulse is completely described by the evolution of $\ket{\psi(0)} = \ket{10}+\ket{11}$ under the Hamiltonian
\begin{equation}
    H_{01}+H_{11} = \left( \begin{array}{cccc}
         0&\frac{\Omega}{2} & 0 & 0  \\
         \frac{\Omega^*}{2}&0 & 0 & 0 \\
         0 & 0 & 0 & \frac{\sqrt{2}\Omega}{2} \\
         0 & 0 & \frac{\sqrt{2}\Omega^*}{2} & 0
    \end{array}\right)
\end{equation}
where the matrix representation is in the $\ket{01}, \ket{0r}, \ket{11}, \ket{W}$ basis. Since the $\ket{rr}$ state can never be populated, it is omitted from the description.

The time-optimal pulse must have maximal amplitude $|\Omega(t)| = \omax$ for all $t$, because if at some time it had sub-maximal amplitude, the pulse could be sped up by simply increasing $\Omega$ and shortening the pulse accordingly. Since the Hamiltonian is proportional to $|\Omega|$, this operation only speeds up the gate, but does not change the trajectory of $\ket{\psi(t)}$. Therefore, the Ansatz $\Omega(t) = \omax e^{i\phi(t)}$ is made. To find the time-optimal $\phi(t)$ we start by fixing a pulse duration $T$ and use GRAPE to minimize the gate error $1-F(\ket{\psi(T)},\theta)$, as given in Eq. \eqref{eq:fidelity_cz}, over the laser phase $\phi(t)$ and the single qubit rotation angle $\theta$. GRAPE is initialized with a random guess for $\phi(t)$ and $\theta$.

The minimal gate error $1-F$ computed by GRAPE for for dimensionless pulse durations $T\omax$ between 0 and $10$ is shown by blue circles in Fig.~\ref{fig:level_scheme_and_cz_panel}b). At $T=0$ the gate error is 0.4, corresponding to the identity gate with $\theta=\pi/2$. As $T$ is increased, the minimal gate error drops and reaches 0 around $T\omax\sim 7.6$. We find that for $T\omax \lesssim 2$ GRAPE can converge to two different values of $1-F$, one of them being the sub-optimal gate error  0.4 -- corresponding to the identity. In the interesting range $T\omax>2$, however, a unique value of $1-F$ is found for all runs of the algorithm. To precisely determine the duration $T_*$ of the time-optimal pulse we plot the gate errors in the range $7.5 \leq T\omax\leq 7.7$ in log scale in Fig.~\ref{fig:level_scheme_and_cz_panel}(c). The gate errors drop until they are of order of $1-F\lesssim 10^{-10}$, which corresponds to the chosen convergence criterion of the optimization. We fit the gate errors with
\begin{equation}
    1-F = \begin{cases} A(T_*-T)^2 & \text{ if } T<T_* \\ 0 & \text{ if } T\geq T_* \end{cases}
\end{equation}
and extract the time-optimal pulse duration $T_*\omax = 7.612$ for the CZ gate. The form of the fitting function is motivated by the observation that $1-F$ should vanish in the first order of $T-T_*$, otherwise negative gate errors would be possible for $T>T_*$. The resulting time-optimal pulse is about 10\% faster then the pulse in Ref.~\cite{levine_parallel_2019} with $T\omax = 8.58536$, showing a relevant speedup of the gate. Our results also show that the duration of the pulse in Ref.~\cite{levine_parallel_2019} is already quite close to the achievable minimum.

The laser phase $\phi(t)$ that leads to the time-optimal pulse as obtained by the GRAPE algorithm is shown as a solid blue line in Fig.~\ref{fig:level_scheme_and_cz_panel}(d) as a function of the dimensionless time $t\omax$. The shape of $\phi$ is the piecewise constant approximation of a smooth curve, which resembles, but is not identical to, a sine wave with a linear offset: It first increases  to $\phi\simeq 1.0$ (at $t\omax\simeq 2.4$ ), then  decreases to $\phi\simeq-0.4$ (at $t\omax\simeq5.2$), and finally  increases again to $\phi=0.7$. We show below in Sec.~\ref{sec:semi-analytical_description_using_the_pmp} using the PMP that the piecewise constant curve found by GRAPE is in fact the discrete approximation of a ``true'' time-optimal pulse, which is smooth [dashed black curve in Fig.~\ref{fig:level_scheme_and_cz_panel}(d)]. Note that our pulse is significantly simpler, and thus easier to implement, than the pulses found in Refs. \cite{muller_optimizing_2011, goerz_robustness_2014}, which also apply optimal control techniques. Our pulse is also more than three times faster than the pulse found in Ref.~\cite{fu_high-fidelity_2022} by optimizing the amplitude of $\Omega$ while assuming a fixed laser detuning. We attribute this to our approach of including the laser phase in the optimization, and to our use of the simplest possible model of the system, which allows to identify the fundamentally time-optimal pulse regardless of experimental imperfections.

As discussed in Sec.~\ref{subsubsec:symmetry_operations} above, a pulse implementing a CZ gate can also be obtained by complex conjugation and/or time-reversal of the pulse shown in Fig. \ref{fig:level_scheme_and_cz_panel}(d). No sign flip of $B$ is necessary during this symmetry operations, because a pulse implementing a certain phase gate at $B=\infty$ implements the same phase gate at $B=-\infty$. When applying complex conjugation or time reversal, the single qubit rotation angle needs to be changed to $-\theta$. For the time-optimal global pulse for the CZ gate, complex conjugation and time reversal coincide, so there are only two time-optimal global pulses: The pulse in Fig.~\ref{fig:level_scheme_and_cz_panel}(d) and the complex conjugated pulse.

The occupation probabilities of $\ket{0r}$ (or $\ket{r0}$) -- when starting in $\ket{01}$ (or $\ket{10}$) -- and $\ket{W}$ -- when starting in $\ket{11}$ --  as a function of $t \omax$ are shown in the Inset of Fig.~\ref{fig:level_scheme_and_cz_panel}(d) as a blue solid curve and an orange dashed curve, respectively. The figure shows that the populations of $\ket{0r}$ (and $\ket{r0}$) and  $\ket{W}$ have a maximum and a local minimum at $T_*\omax/2$, respectively. This is due to the fact that since the coupling between $\ket{11}$ and $\ket{W}$ is $\sqrt{2}$ times stronger than the coupling between $\ket{01}$ and $\ket{0r}$ the population of $\ket{W}$ changes with a faster rate than that of $\ket{0r}$.

\subsection{CZ Gate with single-site addressability}
\label{subsec:non_parallel_cz_gate}
A natural question is whether the pulse from the previous section can be sped up by allowing single-site addressability of the atoms. Here we find that this is not the case.

In order to identify the time-optimal non-global pulse, we proceed just like for global pulses above, except that now two independent lasers described by complex $\Omega_1(t)$ and $\Omega_2(t)$ are considered: We thus proceed by minimizing the gate error [see Eq. \eqref{eq:fidelity_time_evolution_operator}] over both phase and amplitude of $\Omega_1$ and $\Omega_2$ (not shown).  Now the evolution of $\ket{10}$ could in principle be different from that of $\ket{01}$ and the state $\ket{11}$ could evolve to arbitrary states in the $\{\ket{11}, \ket{1r}$, $\ket{r1}\}$ subspace. However, as shown by the red squares in Fig.~\ref{fig:level_scheme_and_cz_panel}b)c), the minimal gate error is not reduced when allowing for non-global instead of global pulses. Hence also when allowing for single-site addressability, the time-optimal pulse satisfies $\Omega_1(t) = \Omega_2(t) = \Omega(t)$, where $\Omega(t)$ describes the time-optimal global pulse found above. Single-site addressability thus brings no speed advantage for the implementation of a CZ gate. This is one of the main results of the paper.

\begin{figure}
\centering
\includegraphics[width = \linewidth]{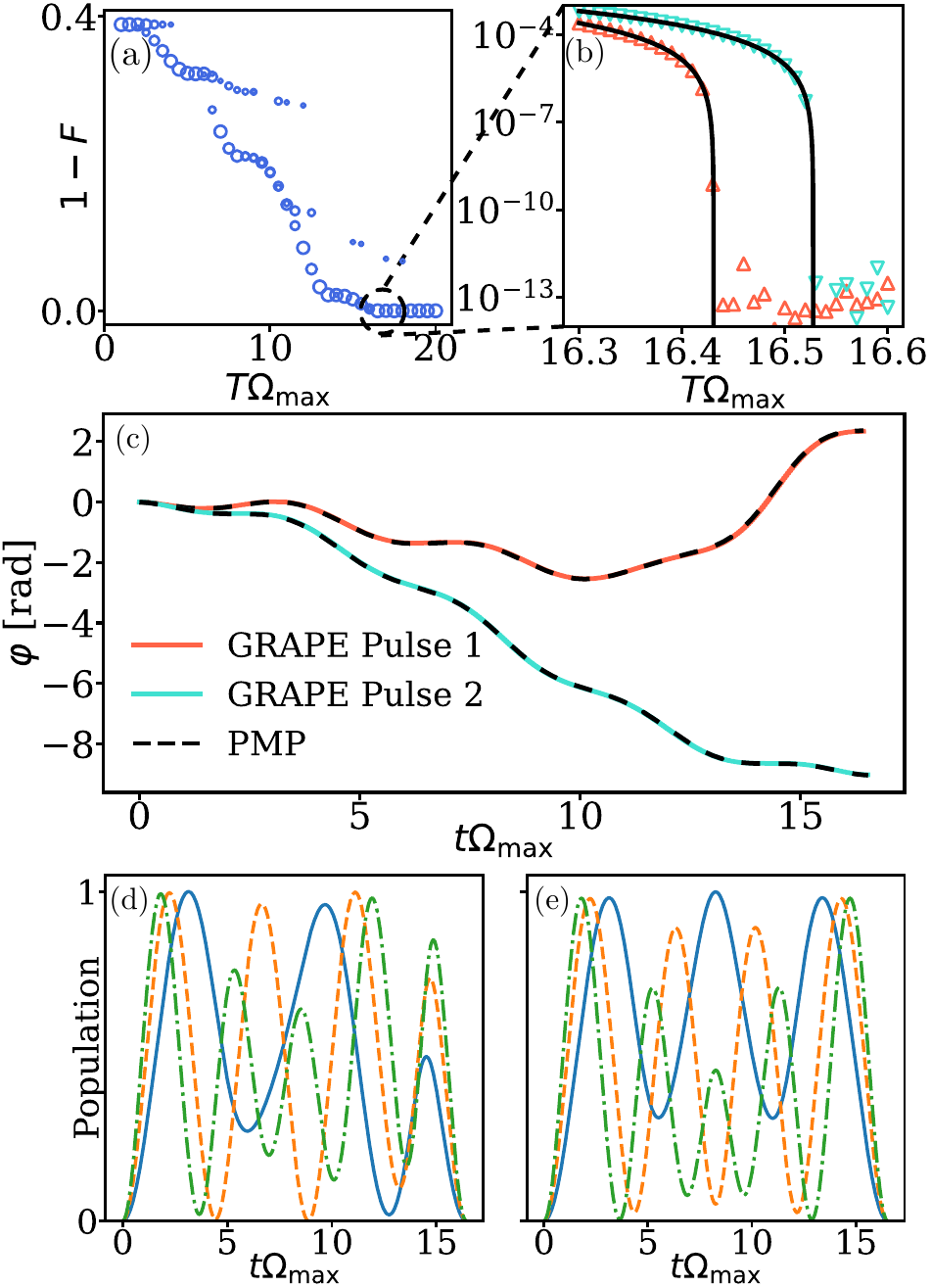}
\caption{\label{fig:ccz}Time-optimal global C$_2$Z gate. a) Smallest possible gate error $1-F$ found by GRAPE at different values of the dimensionless pulse duration $T\omax$. A piecewise constant Ansatz with 399 pieces was used, and at each $T$ GRAPE was started 10 times with a random initialization. The size of the markers is proportional to the number of times that the gate error was within a distance $10^{-5}$ to the position of the marker. b) Zoom-in at the gate errors between $T\omax=16.3$ and $T\omax=16.6$. There are two qualitatively different pulses: Pulse 1 (red, upward pointing triangles) has a duration $T_*^{(1)}\omax =16.43$, Pulse 2 (turquoise, downward pointing triangles) $T_*^{(2)}\omax =16.53$. c) Phase $\phi(t)$ of Pulse 1 and Pulse 2 at their respective $T_*$. Solid lines show the pulse found by GRAPE, dashed lines the fit with the PMP.  d)(e)) Population of the states $\ket{00r}$ (blue, solid line), $\ket{0}\otimes\ket{W}$ (orange, dashed line) and $\ket{W_1}$ (green, dash-dotted line) during pulse 1(2).}
\end{figure}

\subsection{Global C$_2$Z Gate}
\label{subsec:parallel_ccz_gate}
For the global three-qubit C$_2$Z gate we follow  a similar protocol as for the 2-qubit CZ gate. The goal is to implement a phase gate with $\xi_{001} = \theta$, $\xi_{011} = 2\theta$ and $\xi_{111} = 3\theta + \pi$. Up to single qubit rotations of $\theta$ around the z-axis, this implements the gate  $\ket{xyz} \mapsto (-1)^{xyz} \ket{xyz}$. We use GRAPE to minimize the gate error $1-F(\psi(T), \theta)$ with the fidelity given in Eq. \eqref{eq:fidelity_ccz}, the initial state $\ket{\psi(0)} = \ket{001} + \ket{011} + \ket{111}$, and the Hamiltonian $H_{001} + H_{011} + H_{111}$, given as a matrix by
\begin{equation}
 \left( \begin{array}{cccccc}
         0&\frac{\Omega}{2} & 0 & 0 & 0 & 0  \\
         \frac{\Omega^*}{2}&0 & 0 & 0 & 0 & 0 \\
         0 & 0 & 0 & \frac{\sqrt{2}\Omega}{2} & 0 & 0 \\
         0 & 0 & \frac{\sqrt{2}\Omega}{2} & 0 & 0 & 0\\
         0 & 0 & 0 & 0 & 0 & \frac{\sqrt{3}\Omega}{2} \\
         0 & 0 & 0 & 0 & \frac{\sqrt{3}\Omega^*}{2} & 0
    \end{array}\right)
\end{equation}
in the $\ket{001}, \ket{00r}, \ket{011}, \ket{0}\otimes\ket{W}, \ket{111}, \ket{W_1}$ basis.

Figure~\ref{fig:ccz}(a) shows the gate error $1-F$ as a function of the dimensionless pulse duration $T\omax$ in the range $0\leq T\omax\leq 20$, as obtained for the C$_2$Z gate using GRAPE. In the figure, the size of the blue dots is proportional to the number of times that each gate error was obtained in the given set of GRAPE runs. The figure shows that the lowest gate errors form a smooth curve made of larger dots -- corresponding to more frequent convergences of the algorithm -- with several plateaus. Minimal gate errors with values  $1-F \lesssim 10^{-12}$ are reached around $T\omax \simeq 16.5$.  In order to precisely determine the optimal $T_*$, Fig.~\ref{fig:ccz}(b) shows a zoom-in of the results of panel (a) in the interesting range $16.3 \leq T\omax \leq 16.6$, in log-scale. The figure shows the existence of two distinct curves with very similar $T_*$: The first curve, called ``Pulse 1'' from now on, has a $T_*^{(1)}\omax = 16.43$, while the second one, called ``Pulse 2'', has a $T_*^{(2)}\omax = 16.53$. Both pulses are faster then the original non-global implementation with $T\omax = 6\pi$~\cite{isenhower_multibit_2011}. The two pulses result from rather different $\phi(t)$, which are shown in Fig.~\ref{fig:ccz}(c): The laser phase for Pulse 1  oscillates from 0 to -2.6 and then back to about 2.3, while the phase of Pulse 2 is monotonically decreasing from 0 to approximately -9.0. Again equivalent pulses can be obtain by time-reversal and complex conjugation of Pulse 1 and Pulse 2. For Pulse 1 this leads to four different pulses, for Pulse 2 only to two different pulses, since time-reversal and complex conjugation coincide. Figures~\ref{fig:ccz}(d) and (e) show the population during Pulse 1 and Pulse 2, respectively, of $\ket{00r}$, (blue solid line), $\ket{0}\otimes\ket{W} = (\ket{01r}+\ket{0r1})/\sqrt{2}$ (orange dashed line), and $\ket{W_1} = (\ket{11r}+\ket{1r1}+\ket{r11})/\sqrt{3}$ (green dash-dotted line). The three types of populations are found to oscillate approximately with the frequencies $\omax$,  $\sqrt{2}\omax$ and $\sqrt{3}\omax$ that are expected if the laser phase is kept constant. Populations in Pulse 2 are further found to be symmetric with respect to time-reversal.

This concludes Sec.~\ref{sec:time_optimal_gates_at_infinite_blockade_strengths}, in which GRAPE was applied to find the time-optimal global pulses for the CZ and the C$_2$Z gate. For both gates the time-optimal pulse is faster then the traditional non-global pulse, for the CZ gate additionally $\sim \!10$\% faster than the pulse in Ref. \cite{levine_parallel_2019}. All pulses are, up to the discretization introduced by GRAPE, simple and smooth pulses. For the CZ gate we additionally found that allowing for individual addressability of the atoms brings no speedup of the gate.

\section{Semi-Analytical Description of the Pulses using the PMP}
\label{sec:semi-analytical_description_using_the_pmp}

\begin{table*}
\centering
\begin{tabular}{|c||c|c|c|}
\hline
& CZ Pulse & C$_2$Z Pulse 1 & C$_2$Z Pulse 2\\ \hline \hline
$\braket{0|\chi_1(0)}$ & 0.11212523i & 0.23064809i & 0.18360502i \\ \hline
\multirow{2}{*}{$\braket{1|\chi_1(0)}$} & -0.70546052 & 0.06339133 & 0.11567664\\
&-0.21533291i&+0.42859775i&+0.46913157i\\\hline
$\braket{0|\chi_2(0)}$ & -0.06056999i & -0.55488096i & -0.54588492i \\ \hline
\multirow{2}{*}{$\braket{1|\chi_2(0)}$} & 0.51526141 & 0.18981665  & 0.14734489\\
& -0.41739917i& -0.03941303i& -0.04548695i\\ \hline
$\braket{0|\chi_3(0)}$ & & 0.29436027i & 0.30789524i \\ \hline
\multirow{2}{*}{$\braket{1|\chi_3(0)}$} & & -0.19206657  & -0.18760792 \\
& &-0.53858404i &-0.53014859i \\ \hline
$T\omax$ & 7.6114828 & 16.426439 & 16.532211 \\ \hline
$1-F$ & $3.1\cdot 10^{-10}$& $3.1\cdot 10^{-7}$& $2.8\cdot 10^{-6}$ \\ \hline
\end{tabular}
\caption{Initial costates, gate durations and gate errors for the CZ gate and for both pulses of the C$_2$Z gate}
\label{tab:pmp_costates}
\end{table*}

The pulses found by GRAPE are  piecewise constant functions obtained using 99 and 399 parameters for the CZ and the C$_2$Z gate, respectively. In the following, the PMP is used to demonstrate that the latter functions are in fact discrete approximations of smooth pulses that are solutions of ordinary differential equations with just 4 and 6 parameters for the CZ and the C$_2$Z gates, respectively.

In order to apply the PMP to the time-optimal global CZ and C$_2$Z gates at $B=\infty$, the maximization condition \eqref{eq:pmp_phase_gates} has to be formulated for $H_{01}, H_{11}$ and $H_{001}, H_{011}, H_{111}$ as a function of the laser phase $\phi$. The Hamiltonians are, in the appropriate basis,
\begin{align}
   H_q = \frac{\omax}{2}\sqrt{m_q}(\cos\phi \sigma_x - \sin\phi\sigma_y)
\end{align}
with $m_q = \sum_i q_i$ the number of 1s in $q$ and $\sigma_x$ and $\sigma_y$ the Pauli X- and Y-matrices, respectively. The maximization condition \eqref{eq:pmp_phase_gates} is
\begin{equation}
    \phi(t) = \underset{\phi'}{\mathrm{argmax}} \sum_q \Im(\braket{\chi_q(t)| H_q(\phi') |\psi_q(t)}).
    \label{eq:pmp_maximization_condition}
\end{equation}
with $q$ summed over $\{01, 11\}$ for the CZ gate and $\{001, 011, 111\}$ for the C$_2$Z gate. Note that contrary to many other applications of the PMP to quantum optimal control \cite{lin_application_2019, lin_time-optimal_2020, bao_optimal_2018, yang_optimizing_2017, garon_time-optimal_2013}, in our case the Hamiltonian depends on the control $\phi$ in a non-linear way. Therefore, the optimal control is not given by a so-called  ``bang-bang'' pulse, but by a continuous function $\phi(t)$.

The maximization in Eq.~\eqref{eq:pmp_maximization_condition} then leads to
\begin{equation}
    \cos\phi = A/\sqrt{A^2+B^2} \qquad \sin\phi = -B/\sqrt{A^2+B^2}
    \label{eq:pmp_optimal_phase}
\end{equation}
with
\begin{align}
   A &= \Im \left( \sum_q\sqrt{m_q}\braket{\chi_q|\sigma_x|\psi_q}\right)\\
   B &=  \Im \left( \sum_q\sqrt{m_q}\braket{\chi_q|\sigma_y|\psi_q}\right).
\end{align}
Given the initial costates $\ket{\chi_q(0)}$, the whole pulse can be determined from Eq.~\eqref{eq:pmp_optimal_phase} and the Schrödinger equation for states and costates, as long as the quantity $A^2+B^2$ does not vanish during the pulse. The time-optimal pulse is now found by minimizing the gate error calculated from the final states $\ket{\psi_q(T)}$ over the initial costates $\ket{\chi_q(0)}$ and the pulse duration $T$. This reduces the control landscape from an infinite dimensional space to a 4-dimensional space for the CZ gate ($\ket{\chi_{01}}$ and $\ket{\chi_{11}}$ are both two-dimensional vectors) and a 6-dimensional space for the C$_2$Z gate. Since the $ \ket{\psi_q}$ are constrained on the $\braket{\psi_q|\psi_q}=1$ manifold, the costates can be restricted to the tangent space of this manifold, so they can be chosen such that $\Re(\braket{\chi_q(t)|\psi_q(t)})=0$ for all $t$.

In Ref. \cite{van_damme_robust_2017}, optimal pulses for the robust control of a two-level system haven been  successfully obtained by the use of a gradient descent algorithm in the reduced control landscape. In our case however, such an optimization proved to be unsuccessful without a good initial guess for the $\ket{\chi_q(0)}$,  because small variations in the initial costates can lead to large variations in the final states. Therefore, we introduce the following new procedure to obtain a good initial guess of the $\ket{\chi_q(0)}$ based on the pulses found by GRAPE: Denote by $U_q(t)$ the time evolution operator under $H_q$, so that $\ket{\chi_q(t)} = U_q(t)\ket{\chi_q(0)}$. Equation~\eqref{eq:pmp_maximization_condition}  then implies that for all $t$
\begin{align}
    0 &= \frac{\d }{\d\phi}\sum_q \Im(\braket{\chi_q(t)| H_q(\phi) |\psi_q(t)})\\ \nonumber
      &= \sum_q \Im(\braket{\chi_q(0)|\alpha_q(t)})
\end{align}
with
\begin{equation}
    \ket{\alpha_q(t)} = U_q^\dag(t)\frac{\d}{\d \phi}H_q(\phi(t))\ket{\psi_q(t)}.
\end{equation}
GRAPE gives the states, controls and time-evolution operators at discrete times $t_1,...,t_N$. Letting $A_q = \sum_j \ket{j}\bra{\alpha_q(t_j)}$ we obtain
\begin{equation}
    \sum_q \Im(A_q\ket{\chi_k(0)}) = 0.
    \label{eq:pmp_eq_for_initial_costates}
\end{equation}

 Now the real and imaginary parts of all $A_q$ are combined into a single real matrix $A$, and the real and imaginary parts of all $\ket{\chi_q}$ into a single real vector $v$, such that Eq. \eqref{eq:pmp_eq_for_initial_costates} just becomes $Av = 0$. To find approximate, nonzero solutions to this linear equation $A$ is decomposed as $A = \sum_i \lambda_i w_iv_i^T$. An approximate solution to $Av = 0$, and thus a good initial guess for $\ket{\chi_q(0)}$, is given either by $v_i$ or by $-v_i$, where $v_i$ is the singular vector with the smallest, approximately zero, singular value $\lambda_i$.

Using this approximate solution as a starting point of the minimization of the gate error over the initial costates $\ket{\chi_k(0)}$ and the pulse duration $T$ using the Powell method~\cite{jones_scipy_2001} the optimized values in Table~\ref{tab:pmp_costates} are obtained. Using Eq.~\eqref{eq:pmp_maximization_condition} together with the Schrödinger equation for the states and costates then allows one to reconstruct the pulses found by GRAPE directly from the initial costates. We find that these pulses reconstructed by the PMP are in excellent agreement with the pulses found by GRAPE, as can be seen from Fig.~\ref{fig:level_scheme_and_cz_panel}(d) and Fig.~\ref{fig:ccz}(c). These results demonstrate that the pulses found by GRAPE are simply discrete approximations of smooth time-optimal pulses, which can be obtained by just a few variational parameters. This procedure allows for immediate reproducibility of the time-optimal pulses using the few parameters in Table~\ref{tab:pmp_costates} and Eq.~\eqref{eq:pmp_maximization_condition} together with the Schrödinger equation for the states and costates only.\\

\begin{figure}
\centering
\includegraphics[width=\linewidth]{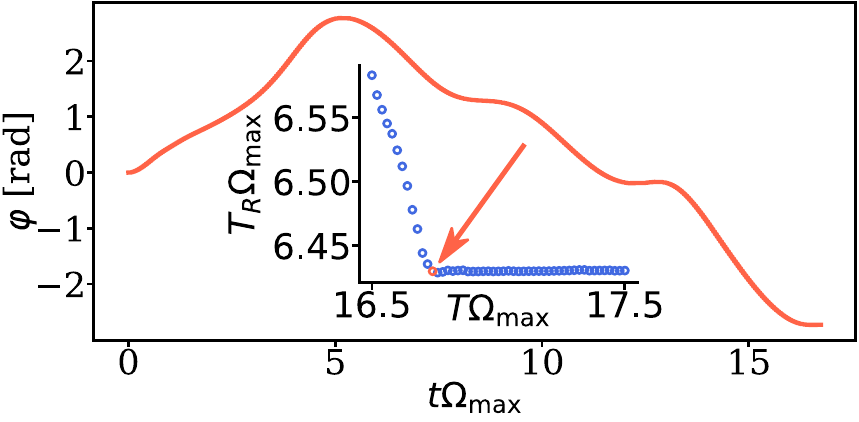}
\caption{\label{fig:decay} Phase of the C$_2$Z pulse at $T\Omega_{\mathrm{max}}=16.74$ where the atoms spend the smallest average time $T_R$ in the Rydberg  state. Inset: Smallest possible value of $T_R\omax$ for different values of the dimensionless gate duration $T\omax$. The duration $T\Omega_{\mathrm{max}}=16.74$ of the shown pulse is marked by the red arrow.}
\end{figure}

\section{Minimizing the Decay of the Rydberg State}
\label{sec:minimizing_the_decay_of_the_rydberg_state}
In the following, we focus on obtaining global pulses for the CZ and C$_2$Z gates that achieve the highest fidelity in the presence of a finite lifetime of the Rydberg state. For this a non-Hermitian term $-\frac{i}{2}\Gamma \ket{r}\bra{r}$ is added to the Hamiltonian of each atom, where $\Gamma$ is the decay rate of the chosen Rydberg state $\ket{r}$. This term describes the loss of population from the $\{\ket{0}, \ket{1}, \ket{r}\}$ subspace of the Hilbert space and thus slightly overestimates the gate error, since it neglects the fact that spontaneous decay of the Rydberg state may repopulate the $\ket{0}$ or $\ket{1}$ state \cite{petrosyan_high-fidelity_2017}. As shown in Appendix~\ref{appendix:infidelity_due_to_decay}, the gate error of a pulse that leads to an exact CZ or C$_2$Z in the decay-free case is now given by $1-F = \Gamma T_R$ for $\Gamma T_R \ll 1$. Here, $T_R$ is the average time that the atoms spend in the Rydberg state, as given by
\begin{equation}
    T_R = \frac{1}{2^n}\sum_{q \in \{0,1\}^n} \int_0^T \braket{\psi_q(t)|\Pi_r|\psi_q(t)} \d t.
    \label{eq:average_time_in_rydberg_state_definition}
\end{equation}
where $\ket{\psi_q(t)} = U(t)\ket{q}$  and $\Pi_r$ is the projector onto the subspace with one atom in the Rydberg state. Our goal is to minimize $T_R$ over all pulses that implement a CZ or C$_2$Z gate with fidelity 1 in the decay-free case. For this, $\Gamma \ll \omax$ is chosen and $1-F$ minimized using GRAPE. Because $\Gamma \ll \omax$ the resulting pulses still have a fidelity $F\approx 1$ in the decay-free case. Because $\Gamma > 0$ the resulting pulses minimize $T_R$ over all pulses with fidelity 1 in the decay free case.

To use GRAPE we fix $\Gamma/\omax= 10^{-4}$, but we stress that the resulting pulse minimizes $T_R$ for any $\Gamma$ and thus minimizes the gate error as long as $\Gamma T_R \ll 1$.  For the CZ gate we then find that the time-optimal pulse is essentially indistinguishable from the pulse with the lowest $T_R$: $T_R\omax$ takes the values $T_R\omax=2.957$ and $T_R\omax=2.947$ for the former and the latter cases, respectively, where the latter is evaluated for a comparatively long pulse duration $T\omax=30$. On the contrary, for the C$_2$Z gate the minimal $T_R$ is improved with respect to both Pulses 1 and 2 found above (which have $T_R\omax = 6.90$ and $T_R\omax = 7.52$, respectively). This is shown in the Inset of Fig.~\ref{fig:decay}, which displays $T_R$ for the pulses leading to the lowest gate errors in the range $16.5 <T\omax < 17.5$: $T_R$ is found to sharply decrease to a value $T_R\omax \simeq 6.431$ at $T\omax\simeq16.74$ and then to stay constant for larger $T$. We have checked that this behavior persists for comparatively long times of at least $T\omax = 30$ -- twice as long as the time-optimal pulse.  The pulse improves $T_R$ by about 7\% and 14\% over the $T_R$ of the time-optimal Pulses 1 and 2, respectively, clearly demonstrating that in general the time-optimal pulse is not identical to the pulse with the smallest average time in the Rydberg state for the three-qubit C$_2$Z gate. Consistently, Fig.~\ref{fig:decay} shows that the phase of the pulse that minimizes $T_R$ at $T\omax = 16.74$ is qualitatively different from that of both time-optimal Pulses 1 and 2 found above (see Fig.~\ref{fig:ccz}(c)).

These results demonstrate that GRAPE can also be used to find the pulse minimizing the time spent in the Rydberg state, instead of the total pulse duration. While for the CZ gate the time-optimal pulse coincides with the pulse minimizing $T_R$, for the C$_2$Z gate a slight reduction of $T_R$ compared to the time-optimal pulse can be achieved.

\section{Gates at Finite Blockade Strength}
\label{sec:gates_at_finite_blockade_strength}
The assumption $B=\infty$ can never be achieved in a real experiment, so it is a  natural question to ask how pulses can operate at finite $B$. There are at least two approaches to this problem: In the first approach, treated in Sec.~\ref{subsec:fixed_blockade_strength}, $B$ is assumed to take a fixed value, and global pulses are optimized to achieve fidelity $F=1$ only at this specific value of $B$. However, in many experiments $B$ is not known with high precision, because it depends on the distance between the atoms, which cam fluctuate due to thermal motion. For this reason the second approach, treated in Sec.~\ref{subsec:variable_blockade_strength}, aims to find global pulses that achieve fidelity $F=1$ only at $B=\infty$, but are affected as weakly as possible as $B$ is decreased.

\subsection{Fixed Blockade Strength}
\label{subsec:fixed_blockade_strength}
\begin{figure*}
\includegraphics[width=\textwidth]{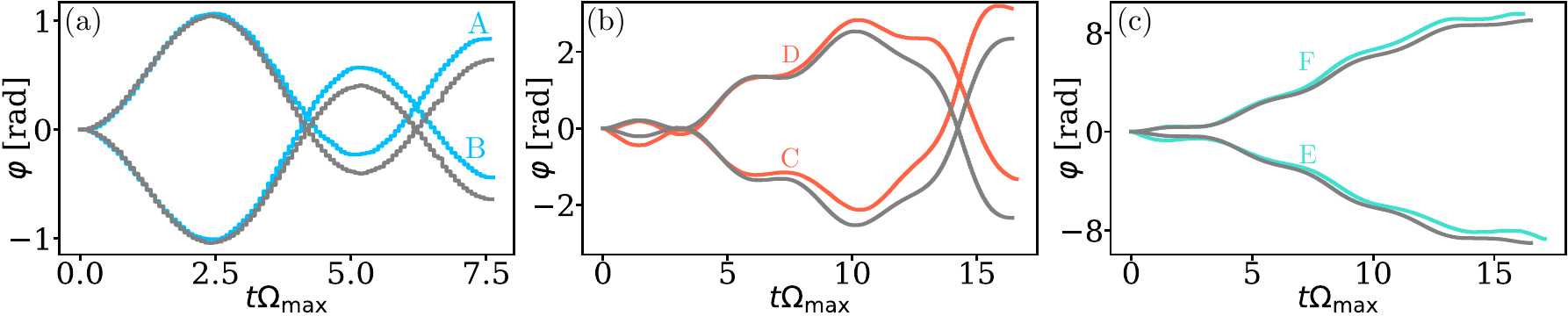}
\caption{\label{fig:finite_blockade1} Pulses with gate error $1-F < 10^{-9}$ at a finite blockade strength $B=10\omax$ (in color) and at $B=\infty$ for comparison (grey). a) For the CZ gate. Pulse A has duration $T_A\omax = 7.58$, pulse B $T_B\omax = 7.64$. b) For the C$_2$Z gate with pulses close to Pulse 1. $T_C\omax = 16.4$, $T_D\omax = 16.6$. c) For the C$_2$Z gate with pulses close to Pulse 2. $T_E\omax = 17.1$, $T_F\omax = 16.2$. }
\end{figure*}
In the first approach, treated in this section, a finite $B$ is fixed and GRAPE is used to find the time-optimal pulses. We begin by discussing CZ gate, where the time-optimal pulses at a finite $B$ are close to the time-optimal pulses at $B=\infty$. Then we find that for the C$_2$Z gate there is in fact a pulse at finite $B$ that is much faster then the time-optimal pulse at $B=\infty$. This pulse significantly populates states with two atoms in the Rydberg state and thus accesses parts of the Hilbert space that are inaccessible in the $B=\infty$ limit. Finally we show that also for the C$_2$Z gate pulses with fidelity 1 at finite $B$ can be found that are close to the time-optimal $B=\infty$ pulses. As a way of example, we take $B=10\omax$ for the rest of this subsection.

\subsubsection{Time-Optimal CZ Gate}
\label{subsubsec:time_optimal_cz_gate}
Just like in Sec.~\ref{subsec:parallel_cz_gate} for $B=\infty$, we now use GRAPE at a finite interaction strength $B=10\omax$. The only modification is that now the $\ket{rr}$ state has to be included, so the pulse is completely describe by the evolution of $\ket{\psi(0)} = \ket{10}+\ket{11}$ under the Hamiltonian
\begin{equation}
    H_{01}+H_{11} = \left( \begin{array}{ccccc}
         0&\frac{\Omega}{2} & 0 & 0 & 0  \\
         \frac{\Omega^*}{2}&0 & 0 & 0 & 0 \\
         0 & 0 & 0 & \frac{\sqrt{2}\Omega}{2} & 0 \\
         0 & 0 & \frac{\sqrt{2}\Omega^*}{2} & 0 & \frac{\sqrt{2}\Omega}{2}\\
         0 & 0 & 0 & \frac{\sqrt{2}\Omega^*}{2} & B
    \end{array}\right)
\end{equation}
in the $\ket{01}, \ket{0r}, \ket{11}, \ket{W}, \ket{rr}$ basis. We find two different pulses with similar pulse duration: The first pulse, ``Pulse A'', with $T_*^{(A)}\omax = 7.574$ qualitatively resembles the  $B=\infty$ pulse from Fig.~\ref{fig:level_scheme_and_cz_panel}(d), while the second, slightly longer, ``Pulse B'' with $T_*^{(B)}\omax = 7.639$ qualitatively resembles the complex conjugated version of the pulse from Fig.~\ref{fig:level_scheme_and_cz_panel}(d). Pulse A and Pulse B are shown together with the two time-optimal pulses in the $B=\infty$ case in Fig.~\ref{fig:finite_blockade1}a.  Pulse A is slightly faster, pulse B slightly slower then the time-optimal pulse for $B=\infty$. Both pulses achieve a gate error below $10^{-10}$, showing that even for a finite $B$ it is possible to implement an exact CZ gate.

\subsubsection{Speeding up the C$_2$Z gate}
\label{subsubsec:speeding_up_the_C2Z_gate}

\begin{figure}[t]
    \centering
    \includegraphics[width=\linewidth]{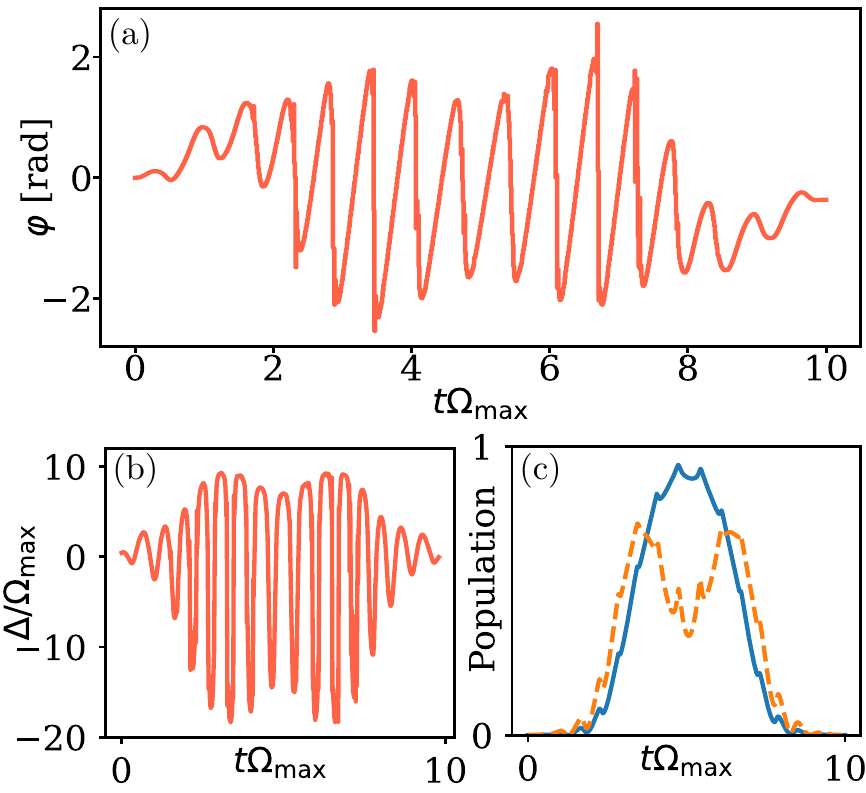}
    \caption{Pulse for a C$_2$Z gate with duration $T\omax = 10$ and gate error $1-F = 2.5\cdot 10^{-10}$  at $B=10\omax$. a) Laser phase $\phi$ over time. b) Detuning $\Delta = \frac{\d \phi}{\d t}$. The detuning is of the order of $B$. c) Population of $\ket{0rr}$ and $\ket{1rr}$ shown as blue solid line and orange dashed line, respectively. Both states with two atoms in the Rydberg state get significantly populated.}
    \label{fig:finite_blockade2}
\end{figure}
For the C$_2$Z gate the effects of a finite blockade strength depend on the positioning of the atoms relative to each other. Here we assume that the blockade strength $B$ is the same for any pair of atoms. In the case of an isotropic blockade strength this can be achieved by placing the atoms on the vertices of an equilateral triangle, the so-called triangular arrangement \cite{barredo_demonstration_2014}. Another possibility is the so-called linear arrangement, in which the three atoms are arranged in a one-dimensional chain, so the distance between the two outer atoms is twice the distance between an outer and the inner atom. We discuss this linear arrangement briefly at the end of this section and in Appendix~\ref{appendix:the_c2z_gate_in_the_linear_arrangement}.

When applying GRAPE to find the time-optimal pulse for a C$_2$Z gate in the triangular arrangement at a finite $B$, pulses with a duration significantly below the duration of the time-optimal pulse in the $B=\infty$ case are found. As an example, at $B=10\omax$ there exists a pulse with duration $T\omax = 10$ and gate error $1-F = 3\cdot 10^{-10}$, which is shown in Fig.~\ref{fig:finite_blockade2}(a). The phase $\phi(t)$ of this pulse varies rapidly with $t$, such that the detuning $\Delta = \d \phi / \d t$ is of the order of $B$ [Fig. \ref{fig:finite_blockade2}(b)]. This suggests that this pulse operates in the so-called Rydberg antiblockade regime \cite{ates_antiblockade_2007} where states with two atoms in the Rydberg state can become significantly populated. This is confirmed by the populations of $\ket{0rr}$ and $\ket{1rr}$, that are shown as a function of $t$ in Fig.\ref{fig:finite_blockade2}(c) as blue solid line and orange dashed line, respectively. The pulse therefore speeds up the C$_2$Z gate by using states with two atoms in the Rydberg state, which is not possible in the $B=\infty$ case. Naturally one would expect such a pulse to be very sensitive to variations of $B$, and indeed when decreasing $B$ by just 10\% the gate error increases drastically to $1-F = 0.59$. At a finite $B$ it is therefore possible to significantly speed up the C$_2$Z gate at the cost of a high sensitivity to variations in the blockade strength. Since the blockade strength depends strongly on the distance of the atoms, this pulse is only feasible if the distance between the atoms is known with high precision.

\subsubsection{Pulses for the C$_2$Z Gate resembling the $B=\infty$ pulses}
\label{subsubsec:pulses_for_the_C2Z_gate_resembling_the_Binf_case}
In order to find pulses for the C$_2$Z gate that more are robust against variations of the blockade strength, we now determine pulses at finite $B$ that are close to the time-optimal $B=\infty$ pulses. For this, we start by the approximation that the effects of a finite $B$ can be completely captured by an ac-Stark shift of the states with one atom in the Rydberg state. This approximation is justified in in Sec.~\ref{subsec:variable_blockade_strength} and in more detail in Appendix~\ref{appendix:ac_stark_shift} through a time dependent Schrieffer-Wolff transformation. Since in this approximation the Hamiltonians are close to the Hamiltonians in the $B=\infty$ case, we also expect the time-optimal pulses to be close to Pulse 1 and Pulse 2. With this approximation, $H_{011}$ and $H_{111}$ are given by
%\begin{equation}
%    H_{011} = \ket{0}\bra{0} \otimes \left(\frac{\sqrt{2}\Omega}{2}\ket{11}\bra{W}+\hc - \frac{|\Omega|^2}{2B}\ket{W}\bra{W}\right)
%    \label{eq:c2Z_hamiltonian_stark1}
%\end{equation}
\begin{align}
  \label{eq:c2Z_hamiltonian_stark1}
     H_{011} = \ket{0}\bra{0} \otimes \Bigg(&\frac{\sqrt{2}\Omega}{2}\ket{11}\bra{W}+\hc \\ \nonumber
     &- \frac{|\Omega|^2}{2B}\ket{W}\bra{W}\Bigg)
\end{align}
and
\begin{equation}
    H_{111} = \frac{\sqrt{3}\Omega}{2}\ket{111}\bra{W_1} + \hc - \frac{|\Omega|^2}{B}\ket{W_1}\bra{W_1}
    \label{eq:c2Z_hamiltonian_stark2}
\end{equation}
GRAPE is used to find the time-optimal pulses under this approximation. These pulses are then used as initial guesses for GRAPE with the exact Hamiltonian. Through this procedure, pulses with fidelity $F=1$ at finite $B$ resembling the $B=\infty$ pulses are found. Pulses C and D, close to Pulse 1 and its complex conjugated version, respectively, are shown Fig.~\ref{fig:finite_blockade2}(b). Their durations are $T^{(C)}\omax = 16.4$ and $T^{(D)}\omax = 16.6$. Pulses E and F, close to Pulse 2 and its complex conjugated version, respectively,  are shown in Fig.~\ref{fig:finite_blockade2}(c), with durations $T^{(E)}\omax = 17.1$ and $T^{(F)}\omax = 16.2$. All pulses have a gate error  $1-F < 10^{-9}$ at $B=10\omax$ They are also significantly less susceptible to variations of $B$ than the faster pulse found above, i.e. reducing $B$ by 10\% never increases the gate error to more then $4\cdot 10^{-4}$. This is an improvement by three orders of magnitude compared to the faster pulse found above.

For the linear instead of the triangular arrangement it is shown in Appendix~\ref{appendix:the_c2z_gate_in_the_linear_arrangement} that, under the approximation that the effects of a finite $B$ can be completely captured by an ac-Stark shift and in the $B\gg\omax$ limit, there exist no pulses with fidelity $F=1$ close to the $B=\infty$ pulses. Qualitatively, this is caused by different pairs of atoms being affected differently by the finite $B$, but in the same way by the global laser.

In summary, the results of this section show that for both the CZ and the C$_2$Z gate in the triangular arrangement the time-optimal pulses at $B=\infty$ can be modified to compensate for a finite value of $B$. For the C$_2$Z gate it is even possible to significantly decrease the gate duration in the finite $B$ case by a pulse which significantly populates states with two atoms in the Rydberg state. This comes at the downside of a much larger sensitivity to variations in $B$. Note that for the CZ gate a similar speedup of the gate is not possible. All results were obtained at $B=10\omax$. We expect that the picture remains qualitatively the same as long as $B\gg \omax$.

\subsection{Variable Blockade Strength}
\label{subsec:variable_blockade_strength}

\begin{figure*}[t]
\includegraphics[width=\textwidth]{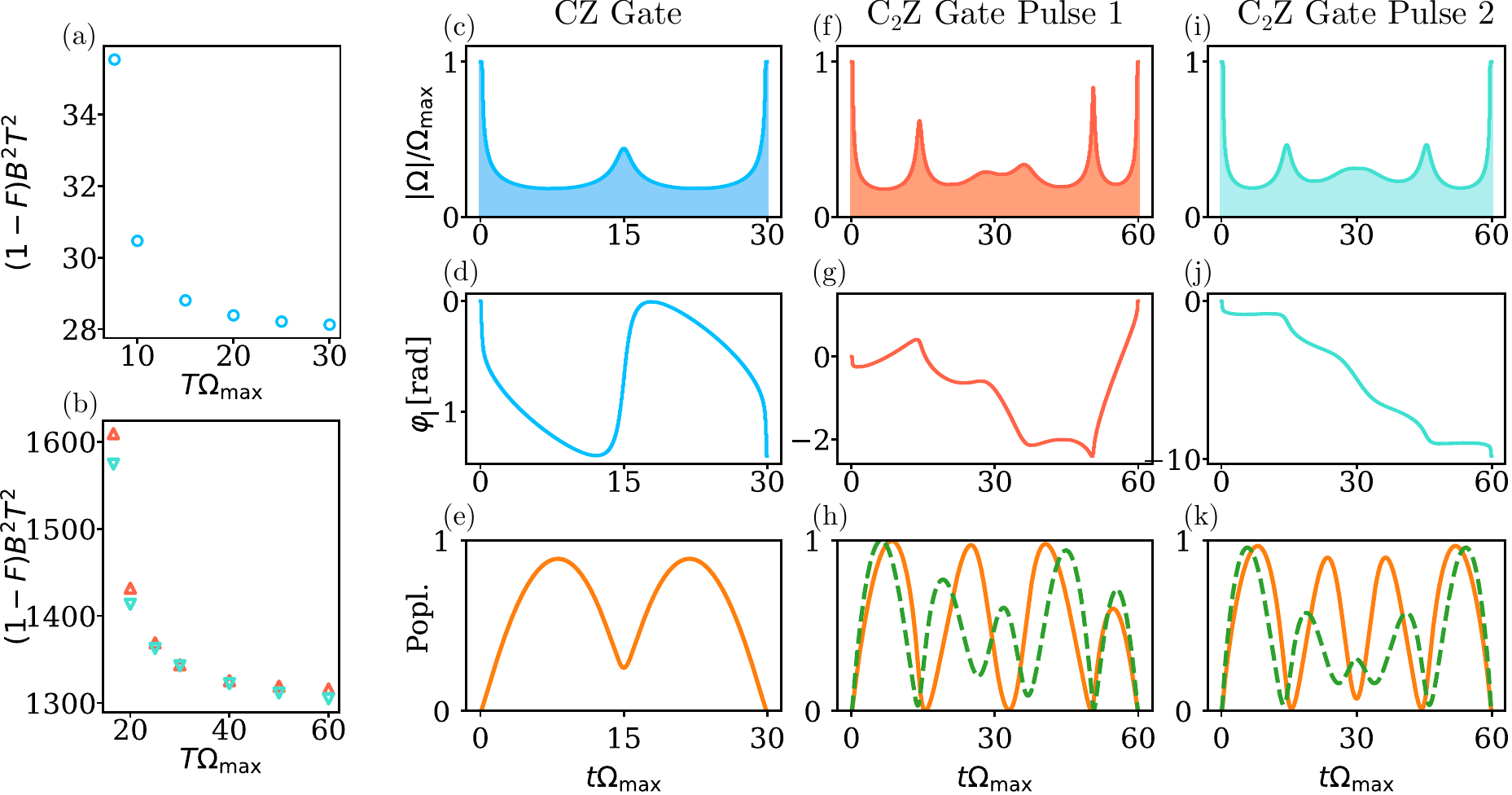}
\caption{\label{fig:blockade_robustness}Pulses for the CZ and the C$_2$Z gate that are as robust as possible against variations of the blockade strength. a) Rescaled gate errors $(1-F)B^2T^2$ for different values of the dimensionless gate duration $T\omax$ for the CZ gate. b) Rescaled gate errors $(1-F)B^2T^2$ for the C$_2$Z gate. Red upward pointing triangles show optimization results when initializing GRAPE with the time-optimal Pulse 1, turquoise downward pointing triangles when initializing with Pulse 2. c) (d)) Laser amplitude (phase) minimizing $(1-F)B^2T^2$ for the CZ gate at $T\omax=30$. e) Population of $\ket{W}$ during the pulse from c)/d) as a function of time. The laser amplitude is large when the population of $\ket{W}$ is small. f) (g)) Laser amplitude(phase) minimizing $(1-F)B^2T^2$ for the C$_2$Z gate when initializing GRAPE with Pulse 1 at $T\omax=60$. h) Population of $\ket{0}\otimes\ket{W}$ (orange solid line) and $\ket{W_1}$ (green dashed line) during Pulse 1 as a function of time. The laser amplitude is large when the populations of $\ket{0}\otimes\ket{W}$ and $\ket{W_1}$ are small. i),j),k) Analogous to f),g),h) for Pulse 2 instead of Pulse 1.}
\end{figure*}

Instead of designing a pulse to work at a specific blockade strength $B$, in this section only pulses that have fidelity $F=1$ at $B=\infty$ are considered. Within the set of these pulses, the pulses that are affected the least by a finite $B$ are identified. To this end, it is shown in Appendix~\ref{appendix:ac_stark_shift} using a time dependent Schrieffer-Wolff transformation \cite{theis_nonadiabatic_2017} that to first order in $1/B$ all effects of the finiteness of $B$ can be described by an ac-Stark shift of the energy of the states with one atom in the Rydberg state. For the CZ gate this means modifying $H_{11}$ to
\begin{equation}
    H_{11} = \underbrace{\frac{\sqrt{2}\Omega}{2}(\ket{11}\bra{W}+\hc)}_{H_{11}^{(0)}} \underbrace{- \frac{|\Omega|^2}{2B}\ket{W}\bra{W}}_{H_{11}^{(1)}/B}
\end{equation}
where we split $H_{11} = H_{11}^{(0)} + \frac{1}{B}H_{11}^{(1)}$. We also expand the state as $\ket{\psi_{11}(t)} =U(t)\ket{11} =  \ket{\psi_{11}^{(0)}(t)} + \frac{1}{B}\ket{\psi_{11}^{(1)}(t)} + \mathcal{O}(B^{-2})$.

As shown in Appendix~\ref{appendix:infidelity_to_second_order_in_B}, a pulse with fidelity $F=1$ in the $B=\infty$ case has at finite $B$ a gate error of
%\begin{equation}
%    (1-F)B^2 = \left( \frac{1}{4} \braket{\psi_{11}^{(1)}(T)|\psi_{11}^{(1)}(T)} - %\frac{1}{10}\left|\braket{0|\psi_{11}^{(1)}(T)}\right|^2 \right). %+ \mathcal{O}(B^{-3]})
%    \label{eq:blockade_robustness_infid}
%\end{equation}
\begin{align}
    \label{eq:blockade_robustness_infid}
     (1-F)B^2 =&\frac{1}{4} \braket{\psi_{11}^{(1)}(T)|\psi_{11}^{(1)}(T)} \\ \nonumber
     &- \frac{1}{10}\left|\braket{11|\psi_{11}^{(1)}(T)}\right|^2  + \mathcal{O}\left(B^{-1}\right)
\end{align}

To lowest order, the gate error thus increases quadratically with $1/B$. Our goal is now to use GRAPE to minimize $\left.\frac{1}{2}\frac{\d^2 (1-F)}{\d (1/B)^2}\right|_{B=\infty}$, as given by the right hand side of Eq.~\eqref{eq:blockade_robustness_infid}. Note that this minimizes the gate error simultaneously for all values of $B$ for which contributions of order $B^{-3}$ and higher can be neglected.

To apply GRAPE, $\ket{\psi_{11}^{(0)}}$ and $\ket{\psi_{11}^{(1)}}$ are treated as independent states satisfying
\begin{equation}
    \frac{\d }{\d t}\left( \begin{array}{c} \ket{\psi_{11}^{(0)}} \\ \ket{\psi_{11}^{(1)}} \end{array}\right) = -i \left( \begin{array}{cc}H_{11}^{(0)} & 0  \\ H_{11}^{(1)}& H_{11}^{(0)} \end{array}\right)\left( \begin{array}{c} \ket{\psi_{11}^{(0)}} \\ \ket{\psi_{11}^{(1)}} \end{array}\right).
    \label{eq:generalized_seq}
\end{equation}
Eq.~\eqref{eq:generalized_seq} now replaces the Schrödinger Equation for $\ket{\psi_{11}}$ in the formulation of GRAPE.

To minimize  $\left.\frac{1}{2}\frac{\d^2 (1-F)}{\d (1/B)^2}\right|_{B=\infty}$ over all states with fidelity $F=1$ in the $B=\infty$ case, the objective function for GRAPE is taken as
\begin{equation}
    J = CF(\ket{\psi_{01}},\ket{\psi_{11}^{(0)}},\theta) + \left.\frac{1}{2}\frac{\d^2 (1-F)}{\d (1/B)^2}\right|_{B=\infty}.
\end{equation}
 Here, $C$ is a large constant ensuring that only pulses with fidelity close to 1 in the $B=\infty$ case can minimize $J$. We take $C=10^4$ and verify that indeed the gate errors at $B=\infty$ are always below $3\cdot 10^{-6}$.

 Given a pulse $\Omega(t)$ with duration $T$ that implements a CZ gate at $B=\infty$, the gate error at finite $B$ can be reduced by an arbitrary factor $\beta^2$, for $0<\beta<1$, by simply stretching the pulse to duration $T/\beta$ and taking $\tilde{\Omega}(t) = \beta\Omega(\beta t)$. To see this, note that $H_{11}^{(1)}$ is decreased by $\beta^2$, while the pulse duration is increased by $1/\beta$. Hence $\ket{\psi_{11}^{(1)}}$ is decreased by $\beta$, and $1-F$ according to Eq. \eqref{eq:blockade_robustness_infid} by $\beta^2$. To compare different pulses beyond a stretch, the dimensionless quantity $\alpha = (1-F)B^2T^2$ is taken as a quality measure for pulses from now on.

Using Eq.~\eqref{eq:blockade_robustness_infid}, the time-optimal pulse for the CZ gate is found to have $\alpha = 35.9$. To improve upon this, GRAPE is now used to minimize $\alpha$ over both the amplitude and the phase of the laser pulse at a fixed $T$, using the time-optimal pulse stretched to duration $T$ as initial guess. The minimal values of $\alpha$ for values of $T\omax$ between 7.61 and 30 are shown in Fig.~\ref{fig:blockade_robustness}(a). $\alpha$ decreases as the pulse duration increases, but asymptotically approaches a non-zero value around $\alpha \approx 28$ as $T\rightarrow \infty$, which is an improvement of over 20\% compared to the time-optimal pulse. The amplitude and the phase of $\Omega(t)$ at $T\omax = 30$ are shown in Figs.~\ref{fig:blockade_robustness}(c) and (d) respectively. The amplitude starts maximal, then drops to about 25\% after a quarter of the pulse duration. Towards the middle of the pulse the amplitude increases again to around 50\%, then it decreases again to 25\% at three quarters of the pulse duration and finally increases to the maximal amplitude at the end of the pulse. This behavior can be understood by considering the population of $\ket{W}$, the only state affected by the ac-Stark shift due to the finite $B$. The population of $\ket{W}$, shown in Fig.~\ref{fig:blockade_robustness}(e), starts at 0 and increases to 0.9 at $t\approx 0.25T_*$. It then decreases to 0.25 at $t=0.5T_*$ and increases again to 0.9 at $t\approx 0.75T_*$, before dropping to 0 at the end of the pulse. Notably, the laser amplitude is inversely correlated to the population $\ket{W}$. Through this, whenever the population of $\ket{W}$ is large, the laser amplitude, and thus also the ac-Stark shift of $\ket{W}$, is reduced.

For the C$_2$Z gate in the triangular arrangement the ac-Stark shift due to the finite $B$ leads to $H_{011}$ and $H_{111}$ as given in Eqs.~\eqref{eq:c2Z_hamiltonian_stark1} and \eqref{eq:c2Z_hamiltonian_stark2}. A formula for the gate error depending on $\ket{\psi_{011}^{(1)}(T)}$ and  $\ket{\psi_{111}^{(1)}(T)}$ analogous to Eq.~\eqref{eq:blockade_robustness_infid} is derived in Appendix~\ref{appendix:infidelity_to_second_order_in_B} [Eq.~\eqref{eq:appendix_d_fidelity_ccz_final}]. Again $\alpha = (1-F)B^2T^2$ is used as a quality measure for different pulses, the time-optimal Pulse 1 has $\alpha = 1850$, the slightly slower Pulse 2 has $\alpha = 1660$.  GRAPE is applied with either Pulse 1 or Pulse 2 as initial guess to minimize $\alpha$ for $T\omax$ between 16.6 and 60, shown in Fig.~\ref{fig:blockade_robustness}(b) with red upward pointing triangles for Pulse 1 as initial guess and turquoise downward pointing triangles for Pulse 2 as initial guess.  For both cases $\alpha$ decreases when $T$ is increased and asymptotically approaches $\alpha \approx 1300$ for both pulses, an improvement of 30\% over the time-optimal Pulse 1 and of 20\% over Pulse 2 . The amplitude and phase for the pulse minimizing $\alpha$ at $T\omax = 60$ when initializing GRAPE with pulse 1 are shown in Figs.~\ref{fig:blockade_robustness}(f) and (g) respectively, the amplitude and phase when initializing GRAPE with Pulse 2 in Figs.~\ref{fig:blockade_robustness}(i) and (j) respectively. The laser amplitude is again inversely correlated to the populations $\ket{0}\otimes\ket{W}$ and $\ket{W_1}$, shown in Fig.~\ref{fig:blockade_robustness}(h) and (k) as orange solid line and green dashed line respectively, and displays several peaks at times where the population of these states is small.

The results in this section show that both for the CZ and for the C$_2$Z gate the time-optimal pulses can be improved to decrease the effect of a finite blockade strength at the cost of a longer pulse duration. The improvement of the gate error goes beyond simply stretching the pulses and is based on a modulation of the laser amplitude to reduce the ac-Stark shift when the states affected by it are populated most.

\section{Gate Errors for a Specific Setup}
\label{sec:infidelity_for_a_specific_setup}

\begin{figure}
    \includegraphics{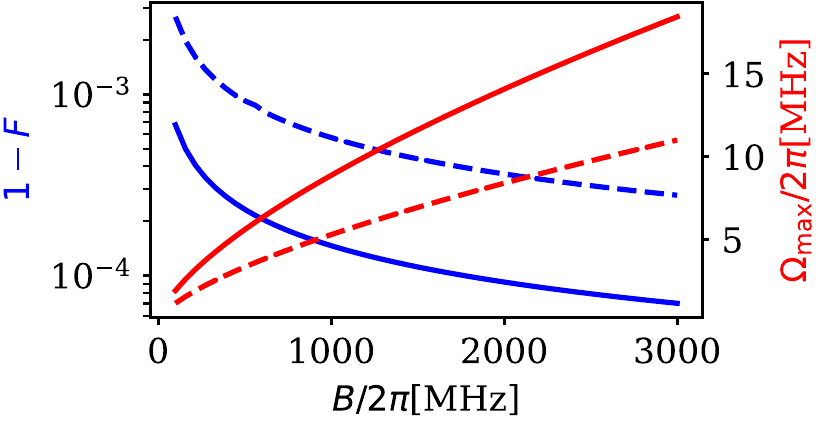}
    \caption{Minimal gate errors (blue, in log scale) and optimal Rabi frequencies $\omax$ (red, in linear scale) for the CZ gate (solid lines) and Pulse 1 of the C$_2$Z gate in the triangular arrangement (dashed lines) for the blockade strength between $100\mathrm{MHz} < B/2\pi < 3\mathrm{GHz}$. The parameters in Eq.~\eqref{eq:specific_infidelity} are $\Gamma=1/540\mu \mathrm{s}$ and $T\omax = 7.612$, $T_R\omax = 2.975$,  $\alpha=35.9$ for the CZ gate and $T\omax = 16.43$, $T_R\omax = 6.9$, $\alpha = 1850$ for Pulse 1 of the C$_2$Z gate.}
    \label{fig:infidelities}
\end{figure}

Finally, in this section we calculate the gate errors of the time-optimal pulses found above at a specific blockade strength and decay rate of the Rydberg state. The same setup as in \cite{saffman_symmetric_2020} is chosen, with the qubits encoded in the clock states of Caesium and the Rydberg state $\ket{r} = \ket{107P_{3/2}\, m_j=3/2}$.  At 300K the Rydberg state has a lifetime of $1/\Gamma = 540\mu\mathrm{s}$. Adding the error contributions from the decay of the Rydberg state and due to a finite blockade strength and neglecting all other error sources, the gate error is given by
\begin{equation}
    1-F = \frac{\Gamma(T_R\omax)}{\omax} + \frac{\omax^2 \alpha}{B^2(T\omax)^2}.
    \label{eq:specific_infidelity}
\end{equation}
The gate error can now be minimized by varying $\omax$ while keeping $T\omax$ and $T_R\omax$ fixed and thus balancing the trade-off between the error due to decay of the Rydberg state and due to a finite blockade strength. The minimal gate errors and the optimal values of $\omax$ for the CZ gate and Pulse 1 for the C$_2$Z gate are shown in Fig.~\ref{fig:infidelities}. At the maximal considered blockade strength $B/2\pi=3\mathrm{GHz}$ the minimal gate errors are $1-F = 7.0\cdot 10^{-5}$ for the CZ and $1-F = 2.8\cdot 10^{-4}$ for Pulse 1 for the C$_2$Z gate in the triangular arrangement. For the blockade strength $B/2\pi = 180$MHz, which is shown to be achievable in Appendix~\ref{appendix:estimation_of_effective_blockade_strength}, the gate errors are $1-F = 4.6\cdot 10^{-4}$ for the CZ gate and $1-F=1.8\cdot 10^{-3}$ for the C$_2$Z gate. The required Rabi frequencies take realistic values of the order of $\omax = 2\pi \times 10\mathrm{MHz}$. For Pulse 2 for the C$_2$Z gate (not shown in Fig.~\ref{fig:infidelities}) the gate error is always slightly above the gate error for Pulse 1, at most by 8\%. Additionally, the optimal Rabi frequency is also always slightly above that for Pulse 1, at most by 7\%.

In summary, these results show that moderate blockade strengths around $B\approx 180$MHz and moderate Rabi frequencies around $\omax \approx 5$MHz are sufficient to achieve gate errors of order of $1-F \lesssim 10^{-3}$ for both the CZ and the C$_2$Z gate. Even lower gate errors can be achieved by higher blockade strengths and larger Rabi frequencies. For the calculations above, the time-optimal pulses from Sec.~\ref{sec:time_optimal_gates_at_infinite_blockade_strengths}, which were optimized at $B=\infty$, were used. For the gates from Sec.~\ref{subsec:fixed_blockade_strength}, which were optimized at a finite $B$, the gate error $1-F =\Gamma T_R$ arises solely due to the decay of the Rydberg state and is inversely proportional to $\omax$. Only the errors due to a finite blockade strength and decay of the Rydberg state were considered in the calculations above.

\section{Conclusion}
\label{sec:conclusion}
In this work the time-optimal global pules for the CZ and the C$_2$Z gate in the blockade regime were identified. The pulse durations of $T\omax=7.612$ for the CZ gate and $T\omax=16.43$ for the C$_2$Z gate improve even upon the traditional, non-global pulses \cite{jaksch_fast_2000, isenhower_multibit_2011} and upon the recently discovered global pulse for the CZ gate with $T\omax = 8.585$ \cite{levine_parallel_2019}. Due to the shortest possible pulse duration, most types of errors which become more detrimental for longer gate durations, like the decay of the Rydberg state or Doppler shifts of the laser frequency, are mitigated. Since the pulses are global, they can be realized by a single laser addressing all atoms simultaneously. No single-site addressability is needed, thus significantly simplifying the experimental requirements. Interestingly, for the CZ gate single-site addressability even brings no speedup over the time-optimal global pulse, showing that single-site addressability, which requires a more complex experimetal setup, is not always advantageous.

The results were obtained using the quantum optimal control techniques GRAPE and PMP, which we combined in a novel way. GRAPE allowed to find the time-optimal pulses in the first place, using several hundreds of variational parameters. The PMP then allowed to reduce the variational parameters to just 4 for the CZ and 6 for the C$_2$Z gate, thus showing that the pulses found by GRAPE are indeed the piecewise constant approximation of a simple smooth pulse given by the solution of an ordinary differential equation. The description by the PMP allows for immediate reproducibility of the time-optimal pulses just from the parameters in Table~\ref{tab:pmp_costates} and without using GRAPE again.

GRAPE was also used to optimize the robustness of pulses for a CZ and C$_2$Z gate against the decay of the Rydberg state and against errors arising due to a finite blockade strength. To mitigate the effects of the decay of the Rydberg state, the average time spent in the Rydberg state was minimized. Interestingly, for the CZ gate the time-optimal pulse coincides with the pulse minimizing the time in the Rydberg state, while for the C$_2$Z gate a small improvement is possible.  Two approaches were taken to mitigate the effects of a finite blockade strength: In the first approach, pulses were optimized at a fixed, finite value of $B$ and it was shown that the time-optimal pulses can be modified to compensate for the finiteness of $B$. In the second approach, pulses were identified that implement a CZ or C$_2$Z gate exactly at infinite blockade strength while minimizing the second derivative of the gate error with respect to the inverse blockade strength. The first approach works best if the blockade strength is known exactly, while the second approach gives a pulse that works well for all large blockade strengths.

Finally, the gate errors were calculated for a plausible experimental setup using the $\ket{107P_{3/2}m_j=3/2}$ state of Caesium as the Rydberg state. Gate errors as low as $1-F=7\cdot 10^{-5}$ for the CZ gate and $1-F=3\cdot 10^{-4}$ for the C$_2$Z gate can be achieved at $B/2\pi=3$GHz and Rabi frequencies of the order of $\omax/2\pi \approx 10$MHz. At the much weaker blockade strength of $B/2\pi=180$MHz still gate errors of $1-F = 4.6\cdot 10^{-4}$ for the CZ gate and $1-F=1.8\cdot 10^{-3}$ for the C$_2$Z gate can be achieved.

The results show that quantum optimal control techniques, especially GRAPE and the PMP, are versatile tools to design gates on Rydberg atoms. In this work, only gates that make use of a single Rydberg state per atom and that operate in the blockade regime were considered. While most gates in current experiments are of this kind, it is known that more robust gates can be achieved when using several Rydberg states per atom \cite{petrosyan_high-fidelity_2017, khazali_fast_2020}, and ultrafast gates can be achieved by operating outside of the Rydberg blockade regime \cite{chew_ultrafast_2021}. We expect that quantum optimal control techniques can also be used to improve these gates.

\section*{Data Availability}
All pulse shapes found in this work are available at Ref.~\cite{pulses_figshare}.

\section*{Acknowledgments}
We are grateful to Shannon Whitlock and Frank Wilhelm-Mauch for stimulating discussions. This research has received funding from the European Union’s Horizon 2020 research and innovation programme under the Marie Skłodowska-Curie grant agreement number 955479. G. P. acknowledges support from the Institut Universitaire de France (IUF) and the University of Strasbourg Institute of Advanced Studies (USIAS).

\onecolumn\newpage
\appendix

\section{Estimation of the Effective Blockade Strength}
\label{appendix:estimation_of_effective_blockade_strength}
\begin{figure*}
    \centering
    \includegraphics{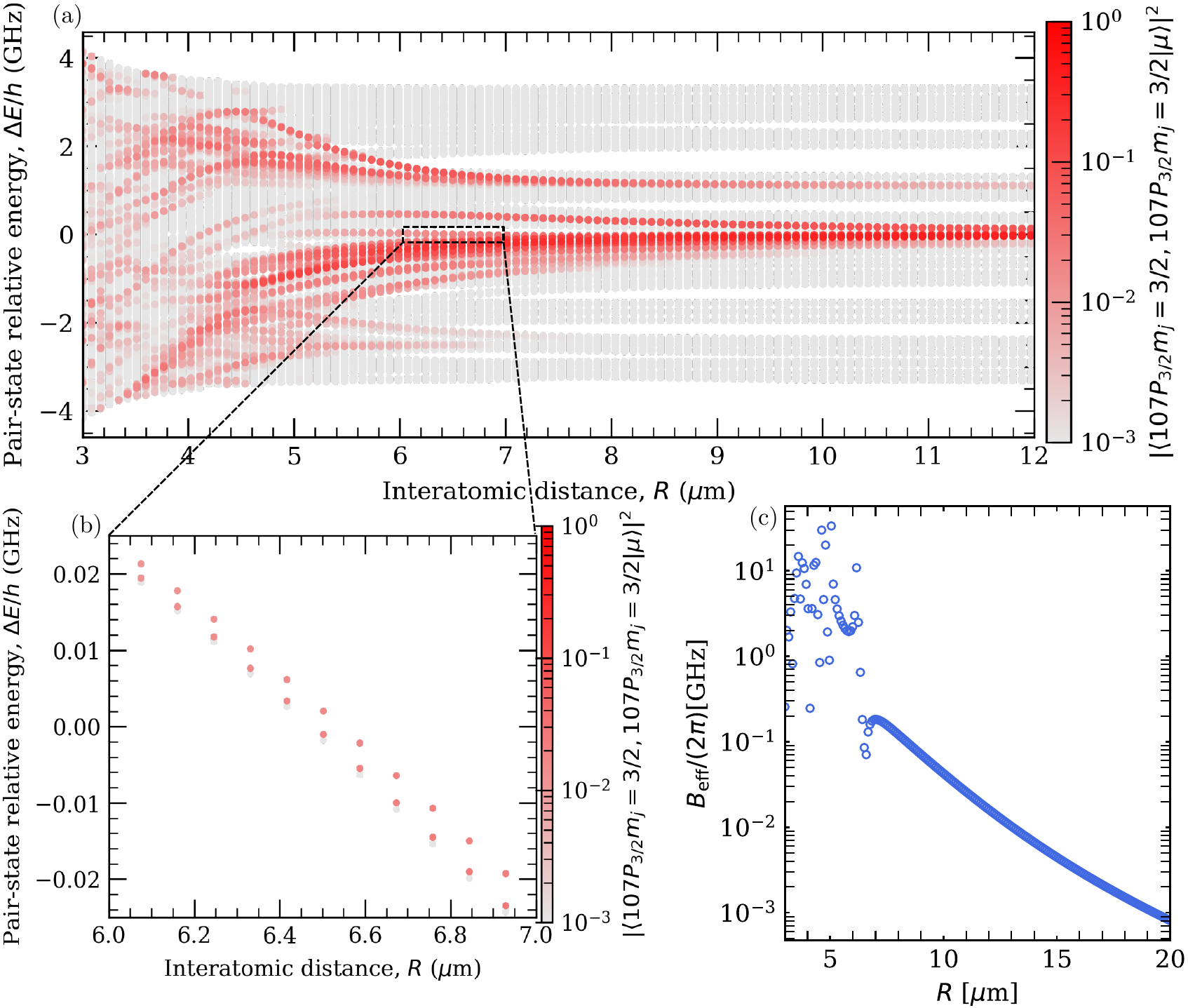}
    \caption{a) Eigenvalues of the Rydberg-Rydberg Hamiltonian $H$ as a function of the interatomic distance $R$. Each marker represents an eigenvalue, the color indicates the squared overlap of the corresponding eigenstate with $\ket{rr}$. b) Zoom-in on the eigenvalues between $6\mu\mathrm{m} < R < 7\mu\mathrm{m}$. There are two eigenstates that become resonant around $R\approx6.5\mu$m, c) The effective blockade strength for different values of $R$.}
    \label{fig:arc_plot}
\end{figure*}
In this appendix we calculate which effective blockade strength can be achieved when using Caesium atoms, $\ket{r} = \ket{107 p_{3/2}\, m =3/2}$ and $\ket{1}=\ket{6S_{1/2}F=4\, m_F=4}$, and aligning the atoms perpendicular to quantization axis. This is the same setup as assumed in \cite{saffman_symmetric_2020, theis_high-fidelity_2016}. The interaction between two atoms at a distance $R$ is given by the dipole-dipole potential
\begin{equation}
    V_{dd} = \frac{e^2}{4\pi\epsilon_0}\frac{\hat{x}_1\cdot \hat{x_2} - 3(\hat{x}_1\cdot \vec{n})(\hat{x}_2\cdot \vec{n})}{R^3}
\end{equation}
where $\hat{x}_i$ denotes the 3d position operator of atom $i$ and $\vec{n}$ is the unit vector along the line interconnecting the atoms. The system of two atoms is then described by the Hamiltonian
\begin{equation}
     H = \sum_{ab} (E_a+E_b)\ket{ab}\bra{ab} + \sum_{abcd} \braket{ab|V_{dd}|cd} \ket{ab}\bra{cd}
\end{equation}
where $a,b,c,d$ denote single atom Rydberg states and $E_a$($E_b$) is the binding energy of the Rydberg state $\ket{a}$($\ket{b}$).

Using the Alkali.ne Rydberg Calculator (ARC) package \cite{sibalic_arc_2017}, $H$ can be easily calculated and diagonalized. To find the effective blockade strength (defined below), we diagonalized $H$, taking into account all Rydberg states with principle quantum number $104 \leq n \leq 110$ and with orbital quantum number $l \leq 3$. The eigenvalues for $R$ between 3$\mu$m and 12$\mu$m are shown in Fig.~\ref{fig:arc_plot}(a), with the color indicating the squared overlap between the corresponding eigenstate and $\ket{rr}$. It can be seen that, contrary to the simple model in Fig.~\ref{fig:level_scheme_and_cz_panel}(a), there are several eigenstates $\ket{\phi_i}$ with a significant overlap with $\ket{rr}$. Since the effects of a finite blockade strength are well approximated through an ac-Stark shift of the $\ket{W}$ state (see Appendix \ref{appendix:ac_stark_shift}), we define the \emph{effective} blockade strength $\beff$ as the blockade strength which correctly reproduces the ac-Stark shift of $\ket{W}$ coming from the coupling to all $\ket{\phi_i}$. This effective blockade strength has been introduced under the name \emph{frequency shift factor} before \cite{walker_consequences_2008}.  $\beff$ can be calculated as
\begin{equation}
    \frac{1}{\beff} = \frac{2}{\Omega^2}\sum_i \frac{|\braket{W|H_{\mathrm{laser}}|\phi_i}|^2}{\Delta_i}
\end{equation}
where $\Delta_i$ is the energy of $\ket{\phi_i}$ (relative to the energy of $\ket{rr}$ in the absence of the dipole-dipole interaction), $H_{\mathrm{laser}}$ describes the laser coupling the $\ket{1}$ state to the Rydberg states and $\Omega$ is the Rabi frequency of this laser for the coupling between $\ket{1}$ and $\ket{r}$.

$\beff$ is shown as a function of $R$ in Fig.~\ref{fig:arc_plot}(c). When decreasing $R$ from infinity, $\beff$ increases monotonically until $R \approx 6.5\mu$m, where it suddenly drops. This drop can be explained by two eigenstates of $H$ becoming resonant ($\Delta_i=0$) around $R \approx 6.5\mu$m, which is shown in Fig.~\ref{fig:arc_plot}(b). These two eigenstates consist primarily of $106S_{1/2} + 109S_{1/2}$ states, with a squared overlap summed over all magnetic quantum numbers of 0.56. The squared overlap with $\ket{rr}$ is 0.02 for both states.

For $R \lesssim 6.5\mu$m the effective blockade strength changes rapidly and non-monotonously with $R$. We therefore take the maximal achievable blockade strength to be the maximal $\beff$ for $R > 6.5\mu$m, which is given by $\beff/2\pi = 180$MHz at $R \approx 7\mu$m.

\section{Gate Error due to Decay}
\label{appendix:infidelity_due_to_decay}
In this appendix, we show that a pulse $\Omega_1,...,\Omega_n$ of duration $T$ which implements a $(\xi_q)_q$ phase gate with fidelity $F=1$ in the absence of decay has a gate error of $1-F = \Gamma T_R$ when a decay of the Rydberg state with decay rate $\Gamma$ with $\Gamma T_R \ll 1$ is included. For this, denote by $H^{(0)}$ the Hamiltonian from Eq.~\eqref{eq:general_hamiltonian} and by $\Delta H = -\frac{i}{2}\Gamma \sum_j \ket{r}_j\leftidx{_j}{\!\bra{r}}{}$ the nonhermitian term added to the Hamiltonian to describe decay. Further, denote by $U^{(0)}(t',t)$ and $U(t',t)$ the time evolution operator from $t'$ to $t$ under $H^{(0)}$ and $H^{(0)}+\Delta H$, respectively. We use the abbreviations $U^{(0)}(t)=U^{(0)}(0,t)$ and $U(t)=U(0,t)$. Then to first order in $\Gamma$ we have
\begin{equation}
    U(T) = U^{(0)}(T) -i \int_0^T \d t U^{(0)}(t,T) \Delta H U^{(0)}(t) =  U^{(0)}(T) -\frac{\Gamma}{2} \int_0^T \d t U^{(0)}(t,T) \sum_j \ket{r}_j\leftidx{_j}{\!\bra{r}}{} U^{(0)}(t).
\end{equation}
Now we use that $U^{(0)}(t,T) = U^{(0)}(T)U^{(0)}(t)^\dag$ and that $\braket{q|U^{(0)}(T)|q} = e^{i\xi_q}$ to obtain
\begin{equation}
    \braket{q|U(T)|q} = e^{i\xi_q}-\frac{\Gamma}{2}e^{i\xi_q} \int_0^T \d t \braket{\psi_q(t)|\Pi_r|\psi_q(t)}
    \label{eq:appendix_infid_decay_braket}
\end{equation}
where $\ket{\psi_q(t)} = U^{(0)}(t)\ket{q}$ and $\Pi_r = \sum_j \ket{r}_j\leftidx{_j}{\!\bra{r}}{}$. Note that for the $B=\infty$ case, where there is at most one atom in the Rydberg state, $\Pi_r$ is the projector onto the states with exactly one atom in the Rydberg state.

Now we insert Eq.~\eqref{eq:appendix_infid_decay_braket} into Eq.~\eqref{eq:fidelity_time_evolution_operator} and expand in first order of $\Gamma$ to obtain
\begin{align}
    F &= \frac{1}{2^n(2^n+1)}\left(  \left|\sum_q e^{-i\xi_q}\braket{q|U(T)|q}\right|^2  + \sum_q \left|\braket{q|U(T)|q}\right|^2\right) \\ \nonumber
    &= \frac{1}{2^n(2^n+1)}\left(  \left|\sum_q 1-\frac{\Gamma}{2} \int_0^T \d t \braket{\psi_q(t)|\Pi_r|\psi_q(t)}\right|^2  + \sum_q \left|1-\frac{\Gamma}{2}\int_0^T \d t \braket{\psi_q(t)|\Pi_r|\psi_q(t)}\right|^2\right) \\ \nonumber
    &= \frac{1}{2^n(2^n+1)} \left(2^{2n}-2^n\sum_q \Gamma \int_0^T \d t \braket{\psi_q(t)|\Pi_r|\psi_q(t)} + 2^n-\sum_q \Gamma \int_0^T \d t \braket{\psi_q(t)|\Pi_r|\psi_q(t)}\right) \\ \nonumber
    &= 1-\Gamma 2^{-n}\sum_q \Gamma \int_0^T \d t \braket{\psi_q(t)|\Pi_r|\psi_q(t)} \\ \nonumber
    &= 1-\Gamma T_R
\end{align}
where in the last equality the definition of $T_R$[Eq.~\eqref{eq:average_time_in_rydberg_state_definition}] was used.

\section{The C$_2$Z Gate in the Linear Arrangement}
\label{appendix:the_c2z_gate_in_the_linear_arrangement}
In this appendix we discuss the C$_2$Z gate in the linear arrangement, in which three atoms are aligned in a row. There, the outer two atoms, labeled atoms 1 and 3 from now on, are separated by a larger distance than an outer and the middle atom (labeled 2). Thus, there are two different blockade strengths: $B_{12} = B_{23} =: B$ and  $B_{13} =: B' < B$. In the van der Waals regime where the blockade strength is proportional to $R^6$, with $R$ the interatomic distance, we have $B' = B/64$, but the following arguments will hold for all $B'$ with  $B' \neq B$.

Let $\Omega^{(0)}$ be a global pulse of duration $T$ that implements a C$_2$Z gate at $B=\infty$, for example Pulse 1 or Pulse 2 from Sec.~\ref{subsec:parallel_ccz_gate}. We will show that if $B,B' \gg \omax$ there is no small change $\Delta\Omega$ such that $\Omega^{(0)}+\Delta\Omega$ implements a C$_2$Z gate at blockade strengths $B, B'$.

Denote by $H_q^{(0)}$ the Hamiltonian under $\Omega^{(0)}$ as defined in Eq.~\eqref{eq:H_q_definition} at $B=\infty$, and by $\Delta H_{q}$ the perturbation of the Hamiltonian due the change $\Delta \Omega$ of the laser pulse and due to the finite Blockade strength. In the limit $B,B' \gg \omax$ we can treat the finite blockade strength through an ac-Stark shift (see Appendix~\ref{appendix:ac_stark_shift}) and obtain
\begin{equation}
    \Delta H_{011}(t) = \frac{\sqrt{2}\Delta\Omega(t)}{2}\ket{011}\bra{W_{011}} + \hc -\frac{|\Omega^{(0)}|^2}{2B}\ket{W_{011}}\bra{W_{011}}
\end{equation}
and
\begin{equation}
    \Delta H_{101}(t) = \frac{\sqrt{2}\Delta\Omega(t)}{2}\ket{101}\bra{W_{101}} + \hc -\frac{|\Omega^{(0)}|^2}{2B'}\ket{W_{101}}\bra{W_{101}}
\end{equation}
with $\ket{W_{011}} = (\ket{01r}+\ket{0r1})/\sqrt{2}$ and $\ket{W_{101}} = (\ket{10r}+\ket{r01})/\sqrt{2}$. Now we denote by $U_q^{(0)}(t',t)$ the time evolution operator from $t'$ to $t$ under $H_q^{(0)}$, set $U_q^{(0)}(t)$ =$U_q^{(0)}(0,t)$   and define $U_q$ analogously with $H_q^{(0)}+\Delta H_q$. Then
\begin{equation}
    U_q(T) = U_q^{(0)}(T) -i \int_0^T \d t U_q^{(0)}(t,T) \Delta H_q U_q^{(0)}(t).
\end{equation}
For the C$_2$Z gate we require $\braket{011|U_{011}(T)|011} = \braket{101|U_{101}(T)|101} = e^{2i\theta}$. We obtain with $U_q^{(0)}(t,T) = U_q^{(0)}(T)U_q^{(0)}(t)^\dag$ that
\begin{equation}
    \braket{011|U_{011}(T)|011} = e^{2i\theta^{(0)}}\left[1-i\left(a_{011} + \frac{b_{011}}{B}\right)\right]
    \label{eq:appendix_B_overlap011}
\end{equation}
and
\begin{equation}
    \braket{101|U_{101}(T)|101} = e^{2i\theta^{(0)}}\left[1-i\left(a_{101} + \frac{b_{101}}{B'}\right)\right]
     \label{eq:appendix_B_overlap101}
\end{equation}
with
\begin{equation}
    a_q = 2\Re\left(\int_0^T \d t \frac{\sqrt{2}\Delta\Omega(t)}{2} \braket{\psi_q(t)|q}\braket{W_q|\psi_q(t)}\right)
\end{equation}
and
\begin{equation}
    b_q = - \int_0^T \d t \frac{|\Omega^{(0)}|^2}{2} |\braket{\psi_q(t)|W_q}|^2
\end{equation}
where again $\ket{\psi_q(t)} = U^{(0)}(t)\ket{q}$.

Because $\ket{\psi_{011}}$ and $\ket{\psi_{101}}$ behave identically up to relabeling the qubits, we obtain $a_{011} = a_{101}$ and $b_{011}=b_{101}$. Since $b_{011}\neq 0$ and $B\neq B'$, it follows from Eqs.~\eqref{eq:appendix_B_overlap011} and \eqref{eq:appendix_B_overlap101} that $\braket{011|U_{011}(T)|011} \neq \braket{101|U_{101}(T)|101}$. Hence the pulse $\Omega^{(0)}+\Delta\Omega$ does not implement a C$_2$Z gate. Qualitatively, this is because the $H_{011}$ and $H_{101}$ are effected differently by a finite blockade strength, but identically by a change in $\Omega(t)$.

\section{Approximation of a Finite Blockade Strength through an AC-Stark Shift}
\label{appendix:ac_stark_shift}

Let $\Omega(t)$ be a global pulse with duration $T$ that implements a CZ or C$_2$Z with fidelity 1 at $B=\infty$. In this appendix we use a time dependent Schrieffer-Wolff transformation (TDSWT) \cite{theis_nonadiabatic_2017} to show that the the final state $\ket{\psi(T)}$ can be obtained to first order in $1/B$ just by including an ac-Stark shift to the Hamiltonian, modifying $H_{11}$ to
\begin{equation}
    \tilde{H}_{11} = \frac{\sqrt{2}\Omega}{2}\ket{11}\bra{W} + \hc - \frac{|\Omega|^2}{2B}\ket{W}\bra{W}
\end{equation}
and $H_{011}$ and $H_{111}$ as in Eqs.~\eqref{eq:c2Z_hamiltonian_stark1} and \eqref{eq:c2Z_hamiltonian_stark2}. We only show this explicitly for the CZ gate, the statement for the C$_2$Z gate follows completely analogously.

The goal of a TDSWT is to make a time dependent basis transformation $e^{-S(t)}$, where $S$ is an anti-hermitian matrix, such that $\ket{\tilde{\psi}_{11}} = e^{S(t)}\ket{\psi_{11}}$ evolves under a Hamiltonian $\tilde{H}_{11}$ that does not couple the $\ket{rr}$ state to the $\ket{W}$ or $\ket{11}$ state anymore. For this, we split  $H_{11} = H_{L1} + H_{L2} + H_{vdW}$ with
\begin{equation}
     H_{L1} = \frac{\sqrt{2}\Omega}{2} \ket{11}\bra{W} +\hc \qquad H_{L2} = \frac{\sqrt{2}\Omega}{2} \ket{W}\bra{rr} + \hc \qquad H_{vdW} = B\ket{rr}\bra{rr}.
\end{equation}

We expand $S=S_1+S_2+...$ and $\tilde{H}_{11} =  \tilde{H}_{11}^{(-1)} + \tilde{H}_{11}^{(0)} + \tilde{H}_{11}^{(1)}+...$ where $S_j$ and  $\tilde{H}_{11}^{(j)}$ are of order $B^{-j}$. There is no $S_0$ term, because in the limit $B=\infty$, $H_{11}$ already does not couple $\ket{rr}$ to $\ket{11}$ or $\ket{W}$. From this limit we can also infer $\tilde{H}_{11}^{(-1)} = H_{vdW}$ and $\tilde{H}_{11}^{(0)} = H_{L1}$. According to the TDSWT, $S_1$ and $\tilde{H}_{11}^{(1)}$ can be found by (see Eqs. (6) and (7) in Ref.~\cite{theis_nonadiabatic_2017}):
\begin{equation}
    [H_{vdW}, S_1] = -H_{L2}
    \label{eq:tsdwt_basis_trafo}
\end{equation}
\begin{equation}
    \tilde{H}_{11}^{(1)} = \frac{1}{2}[H_{L2},S_1].
    \label{eq:tdswt_modified_hamiltonian}
\end{equation}
Solving Eqs.~\eqref{eq:tsdwt_basis_trafo} and \eqref{eq:tdswt_modified_hamiltonian} gives
\begin{equation}
    S_1 = \frac{1}{B}\frac{\sqrt{2}\Omega}{2}\ket{W}\bra{rr} - \hc \qquad  \tilde{H}_{11}^{(1)} = \frac{|\Omega|^2}{2B}\left(-\ket{W}\bra{W} + \ket{rr}\bra{rr}\right)
\end{equation}
We conclude that to first order in $1/B$ we can treat the effects of a finite $B$ just through an ac-Stark shift, as long as we consider $\ket{\tilde{\psi}_{11}}$ instead of $\ket{\psi_{11}}$. However, we are only interested in $\ket{\psi_{11}(t)}$ at $t=0$ and $t=T$, and at both times $\ket{\psi_{11}(t)}$ is, to 0th order in $1/B$, proportional to $\ket{11}$. Because $S_1\ket{11} = 0$ we thus have at these two times $\ket{\tilde{\psi}_{11}(t)} = \ket{\psi_{11}(t)} + \mathcal{O}(B^{-2})$. To describe the effects of a finite $B$ up to first order in $1/B$, it is therefore sufficient to just consider the ac-Stark shift of $\ket{W}$ and use $\tilde{H}_{11}$ instead of $H_{11}$. We remark that terms of order $1/B^2$ are either proportional to $(\Omega/B)^2$ or to $\dot{\Omega}/B^2$. To neglect these terms, $B\gg |\Omega|$ is therefore not sufficient, additionally $B^2 \gg |\dot{\Omega}|$ has to hold. Analogously, to be able to neglect terms of higher orders, $B^n \gg \left| \frac{\d^n}{\d t^n} \Omega \right|$ has to hold for all $n$. If $\Omega$ oscillates with a frequency of the order of $B$, as for example for the pulse in Sec.~\ref{subsubsec:speeding_up_the_C2Z_gate}, this does not hold and the effect of a finite $B$ can not described solely by an ac-Stark shift.

\section{Gate Error of a CZ and C$_2$Z Gate to Second Order in $1/B$}
\label{appendix:infidelity_to_second_order_in_B}
In this appendix we derive perturbative expressions for the gate error at finite $B$ for pulses that implement a CZ or C$_2$Z with fidelity $F=1$ in the $B=\infty$ case. For this, we expand $\ket{\psi_q} = \ket{\psi_q^{(0)}} + \frac{1}{B}\ket{\psi_q^{(1)}} + \frac{1}{B^2}\ket{\psi_q^{(2)}} + \mathcal{O}(B^{-3})$ and show that for the CZ gate it holds that
\begin{equation}
    F = 1-\frac{1}{B^2}\left(\frac{1}{4}\braket{\psi_{11}^{(1)}|\psi_{11}^{(1)}} - \frac{1}{10}\left|\braket{11|\psi_{11}^{(1)}}\right|^2\right) + \mathcal{O}(B^{-3})
    \label{eq:appendix_d_fidelity_cz_final}
\end{equation}
while for the C$_2$Z gate in the triangular arrangement it holds that
\begin{align}
    \label{eq:appendix_d_fidelity_ccz_final}
    F = 1-\frac{1}{72B^2}\Big(&27\braket{\psi_{011}^{(1)}|\psi_{011}^{(1)}} +  9\braket{\psi_{111}^{(1)}|\psi_{111}^{(1)}} -\left|3 \braket{011|\psi_{011}^{(1)}} - e^{-i\theta}\braket{111|\psi_{111}^{(1)}}\right|^2 \\ \nonumber
    -& 3\left|\braket{011|\psi_{011}^{(1)}}\right|^2 - \left|\braket{111|\psi_{111}^{(1)}}\right|^2\Big)
\end{align}

To derive both formulas, we use two ingredients:  Firstly any quantity $x = x^{(0)} + \frac{1}{B} x^{(1)} + \frac{1}{B^2} x^{(2)}+...$ depending on $B$ satisfies
\begin{equation}
    |x|^2 = \left|x^{(0)}\right|^2 + \frac{2}{B}\Re\left((x^{(0)})^*x^{(1)}\right) + \frac{1}{B^2}\left( \left|x^{(1)}\right|^2 + 2\Re\left((x^{(0)})^*x^{(2)}\right)\right) + \mathcal{O}(B^{-3}).
    \label{eq:appendix_d_taylor_expansion}
\end{equation}
Secondly, for any normalized vector $\ket{\phi_q}$ depending on $B$ and with  $\ket{\phi_q^{(0)}}=\ket{q}$ it holds that
\begin{equation}
    1 = \braket{\phi_q|\phi_q} = 1 + \frac{1}{B}\Re\left((\braket{q|\phi_{q}^{(1)}}\right) + \frac{1}{B^2}\left(\braket{\phi_{q}^{(1)}|\phi_{q}^{(1)}} + 2\Re(\braket{q|\phi_{q}^{(2)}})\right)+ \mathcal{O}(B^{-3})
\end{equation}
so that
\begin{equation}
    \Re\left((\braket{q|\phi_{q}^{(1)}}\right) = 0
    \label{eq:appendix_d_cz_normalization1}
\end{equation}
and
\begin{equation}
    \Re(\braket{q|\phi_{q}^{(2)}}) = -\frac{1}{2}\braket{\phi_{q}^{(1)}|\phi_{q}^{(1)}}.
    \label{eq:appendix_d_cz_normalization2}
\end{equation}
For the fidelity of a CZ gate we obtain from Eq.~\eqref{eq:fidelity_cz} with $\ket{\phi_{01}} = e^{-i\theta}\ket{\psi_{01}}$ and $\ket{\phi_{11}} = -e^{-2i\theta}\ket{\psi_{11}}$ that
\begin{equation}
    F = \frac{1}{20}\left(\left| 1 + 2\braket{01|\phi_{01}} + \braket{11|\phi_{11}}\right|^2 + 1 + 2\left|\braket{01|\phi_{01}}\right|^2 + \left|\braket{11|\phi_{11}}\right|^2 \right)
\end{equation}
Now we apply Eq.~\eqref{eq:appendix_d_taylor_expansion}. Because the pulse has fidelity 1 in the $B=\infty$ case we have $\ket{\phi_{01}} = \ket{01}$ and $\ket{\phi_{11}^{(0)}} = \ket{11}$, so that
\begin{align}
    \label{eq:appendix_d_fidelity_cz_expanded}
    F &= \frac{1}{20}\left( |3+\braket{0|\phi_{11}}|^2+3+|\braket{0|\phi_{11}}|^2 \right) \\ \nonumber
      &=  \frac{1}{20}\left( 20 + \frac{10}{B}\Re\left(\braket{0|\phi_{11}^{(1)}}\right) +\frac{1}{B^2}\left(2\left|\braket{0|\phi_{11}^{(1)}}\right|^2+10\Re\left(\braket{0|\phi_{11}^{(2)}}\right)\right)\right) + \mathcal{O}(B^{-3})
\end{align}
By inserting Eqs.~\eqref{eq:appendix_d_cz_normalization1} and \eqref{eq:appendix_d_cz_normalization2} into Eq.~\eqref{eq:appendix_d_fidelity_cz_expanded} we obtain Eq.~\eqref{eq:appendix_d_fidelity_cz_final}.

For the C$_2$Z gate we obtain from Eq.~\eqref{eq:fidelity_ccz} with $\ket{\phi_{001}} = e^{-i\theta}\ket{\psi_{001}}, \ket{\phi_{011}} = e^{-2i\theta}\ket{\psi_{011}}$ and $\ket{\phi_{111}} = -e^{-3i\theta}\ket{\psi_{011}}$ that
\begin{align}
    F = \frac{1}{72}\Big(&\left| 1 + 3\braket{001|\phi_{001}} + 3\braket{011|\phi_{011}} + \braket{111|\phi_{111}}\right|^2  \\ \nonumber
     & +1 + 3\left|\braket{001|\phi_{001}}\right|^2 + 3\left|\braket{011|\phi_{011}}\right|^2 + \left|\braket{111|\phi_{111}}\right|^2 \Big)
\end{align}
We apply Eq.\eqref{eq:appendix_d_taylor_expansion} and use that $\phi_{001} = \ket{001}$, $\ket{\phi_{011}^{(0)}} = \ket{011}$ and $\ket{\phi_{111}^{(0)}} = \ket{111}$ to obtain
\begin{align}
     \label{eq:appendix_d_fidelity_ccz_expanded}
    F = \frac{1}{72}\Bigg[&72 + \frac{1}{B}\Re\left( 54 \braket{011|\phi_{011}^{(1)}} + 18\braket{111|\phi_{111}^{(1)}}\right) +
                     \frac{1}{B^2}\bigg(\left|3 \braket{011|\phi_{011}^{(1)}} + \braket{111|\phi_{111}^{(1)}}\right|^2  \\ \nonumber
                     +& 3\left|\braket{011|\phi_{011}^{(1)}}\right|^2 + \left|\braket{111|\phi_{111}^{(1)}}\right|^2 + \Re\left( 54 \braket{011|\phi_{011}^{(2)}} + 18\braket{111|\phi_{111}^{(2)}}\right)  \bigg)\Bigg]
\end{align}
By inserting Eqs.~\eqref{eq:appendix_d_cz_normalization1} and \eqref{eq:appendix_d_cz_normalization2} into Eq.~\eqref{eq:appendix_d_fidelity_ccz_expanded} we obtain Eq.~\eqref{eq:appendix_d_fidelity_ccz_final}.

\begin{figure*}
    \includegraphics[width=\linewidth]{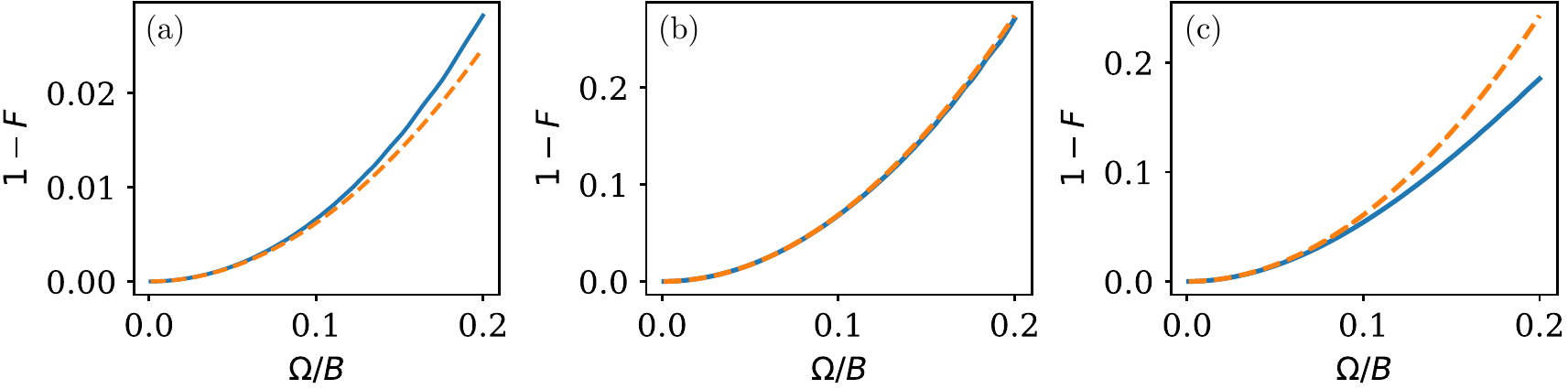}
    \caption{Comparison between the exact gate error at finite $B$ (blue, solid line) and the gate error calculated from Eqs.~\eqref{eq:appendix_d_fidelity_cz_final} and \eqref{eq:appendix_d_fidelity_ccz_final} together with approximating the effects of a finite $B$ through an ac-Stark shift (orange, dashed line). a) For the time-optimal pulse for the CZ gate. b)(c)) For Pulse 1(2) for the C$_2$Z gate. For all three pulses there is an excellent agreement between the exact gate error and the approximation for large enough $B$.}
    \label{fig:infidelity_comparison}
\end{figure*}

We numerically confirm Eqs.~\eqref{eq:appendix_d_fidelity_cz_final} and \eqref{eq:appendix_d_fidelity_ccz_final} as well as the fact that to first order in $1/B$ it is sufficient to account for the finiteness of $B$ through an ac-Stark shift (see Appendix~\ref{appendix:ac_stark_shift}). For this, we consider the time-optimal pulse for the CZ gate and Pulse 1 and Pulse 2 for the C$_2$Z gate (see Sec.~\ref{sec:time_optimal_gates_at_infinite_blockade_strengths}) and calculate the gate error once using the exact Hamiltonian and once from Eqs.~\eqref{eq:appendix_d_fidelity_cz_final} and \eqref{eq:appendix_d_fidelity_ccz_final}. In the latter case, the $\ket{\psi_q^{(1)}}$ are found through Eq.~\eqref{eq:generalized_seq}. The exact and the approximate gate error are shown in Fig.~\ref{fig:infidelity_comparison} as blue solid line and orange dotted line, respectively. For all three considered pulses the exact and the approximate gate error are in excellent agreement for large $B$.

\bibliographystyle{unsrtnat}
\bibliography{bibliography}

\end{document}